\documentclass[12pt]{article}
\usepackage{amsmath,amssymb,amsfonts}
\usepackage{algorithmic}
\usepackage{graphicx}
\usepackage{textcomp}
\usepackage{longtable}
\usepackage{graphicx}
\usepackage{subfigure}
\usepackage{booktabs}
\usepackage[perpage]{footmisc}
\usepackage{setspace}
\usepackage{lscape}
\usepackage{pdflscape}
\usepackage{xcolor}

\begin{document}

\title{Deep Learning in EEG: Advance of the Last Ten-Year Critical Period}
		
\author{Shu~Gong,
		Kaibo~Xing \thanks{S. Gong and K. Xing are with the College of Life Sciences, Sichuan University, Chengdu 610065, China.},
		Andrzej~Cichocki \thanks{A. Cichocki is with the Skolkovo Institute of Science and Technology (SKOLTECH), Moscow 121205, Russia, and also with the Systems Research Institute, Polish Academy of Sciences, 01-447 Warsaw, Poland.},
		and~Junhua~Li \thanks{J. Li is with the School of Computer Science and Electronic Engineering, University of Essex, Colchester CO4 3SQ, UK, and also with the Laboratory
			for Brain-Bionic Intelligence and Computational Neuroscience, Wuyi
			University, Jiangmen 529020, China.} \thanks{This work was supported by the National Natural Science Foundation
	of China under Grant 61806149 and the Guangdong Basic and Applied Basic Research Foundation under Grant 2020A1515010991. This work was also supported by the Ministry  of Education and Science of the Russian Federation under Grant 14.756.31.0001. Corresponding author: Junhua Li (e-mail: juhalee.bcmi@gmail.com)}}
		
\newcommand{\tabincell}[2]{\begin{tabular}{@{}#1@{}}#2\end{tabular}}
\newcommand{\tabincella}[2]{\begin{tabular}{@{}#1@{}}#2\end{tabular}}
	
\maketitle
	
\textbf{Abstract:}

Deep learning has achieved excellent performance in a wide range of domains, especially in speech recognition and computer vision. Relatively less work has been done for EEG, but there is still significant progress attained in the last decade. Due to the lack of a comprehensive and topic widely covered survey for deep learning in EEG, we attempt to summarize recent progress to provide an overview, as well as perspectives for future developments. We first briefly mention the artifacts removal for EEG signal and then introduce deep learning models that have been utilized in EEG processing and classification. Subsequently, the applications of deep learning in EEG are reviewed by categorizing them into groups such as brain-computer interface, disease detection, and emotion recognition. They are followed by the discussion, in which the pros and cons of deep learning are presented and future directions and challenges for deep learning in EEG are proposed. We hope that this paper could serve as a summary of past work for deep learning in EEG and the beginning of further developments and achievements of EEG studies based on deep learning. 


\section{Introduction}

Machine learning technology has benefited to diverse domains in our modern society \cite{jordan2015machine}, \cite{michie1994machine}. Deep learning, a subcategory of machine learning technology, has been showing excellent performance in pattern recognition \cite{lecun2015deep}, dramatically improving classification accuracy. It is worth noting that new world records were created by using deep learning in many competitions such as ImageNet Competition \cite{russakovsky2015imagenet}. The research outcomes of deep learning in speech recognition \cite{amodei2016deep} and computer vision \cite{zhao2019object} have been successfully utilized to develop practical application systems, which are remarkably influencing our life and even changing our lifestyle. 
  
  Deep learning is an enhanced variant of traditional neural network, which is thought to be established based on the inspiration of hierarchical structure existing in visual cortex of the human brain. The adjective 'deep' in the term of deep learning describes the attribute of multiple processing layers forming a long-cascaded architecture. The extracted information becomes more and more abstract from the lowest layer to the highest layer. This is one of the advantages for the deep learning as information expression could be more meaningful when passing onto a higher layer. Meanwhile, deep learning suffers from the issues of slow convergence and high computation demand. These disadvantages have been released by introducing training strategies such as dropout \cite{He2019} and batch normalization \cite{Liu2018}, and the availability of high-performance computers. The high performance is not only due to the capacity improvement of central processing units, but also new computing units such as graphic processing unit and tensor processing unit. These new computing units are designed to suit matrix manipulation, which greatly reduce computational time in deep learning. Moreover, the availability of large scale of data and increased capacity of data storage also promote the use of deep learning.  
	
\begin{figure}
\centering
\includegraphics[width=\textwidth]{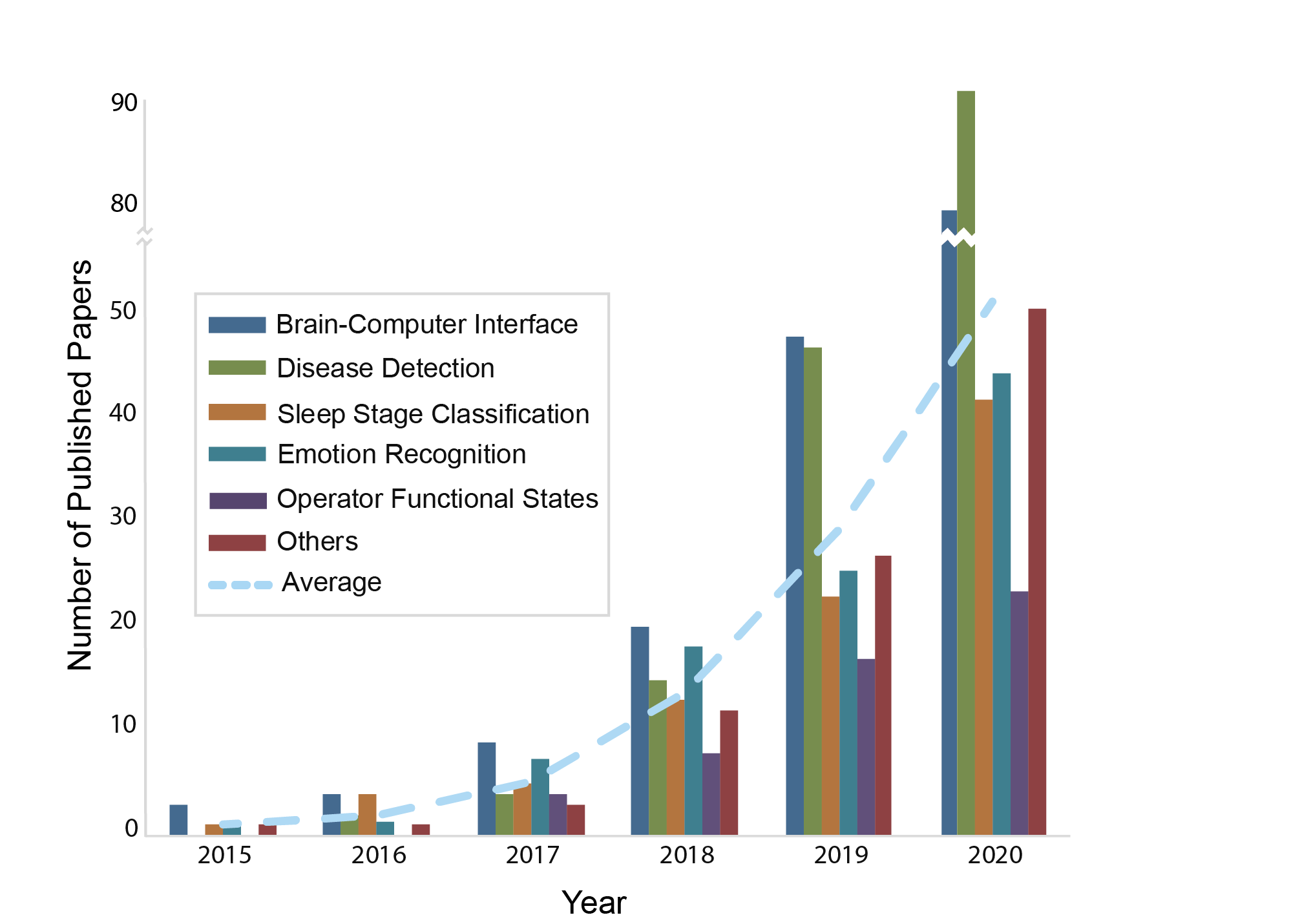}
		\caption{Numbers of the published papers in each year. Note that numbers before 2015 are omitted because of rare papers.}
\label{fig_ratio_year}
\end{figure}
	
	Electroencephalogram (EEG) signal was first recorded by Hans Berger in the year of 1924 \cite{haas2003hans}, which manifests underlying brain activity. Multiple electrodes can be set to record EEG signal by placing them on different locations of the scalp and temporal fluctuations in voltage can be captured in a high resolution (e.g., in milliseconds) by using a high sampling rate. With the advantages of multi-channel recording and high temporal resolution, EEG has been applied to numerous domains from brain-computer interface \cite{wang2006practical, li2012combining, zhang2015sparse, li2015canonical}, to emotion \cite{zheng2014eeg, Chao2019}, to cognition \cite{Li2017}, to brain diseases \cite{Cao2019}. EEG processing methodology is evolved from simple methods such as mean and amplitude comparison to complicated methods such as connectivity topology and deep learning. In particular, deep learning exhibits better performance in EEG classification (a.k.a., recognition or identification) compared to conventional methods (e.g., support vector machine). By using deep learning, discriminative features could be extracted without handcraft, which requires specific knowledge and expertise. It could avoid the low performance derived from unsuitable handcrafted features. However, deep learning is not a destination because model architecture and parameters have to be set manually. A good classification performance is usually not obtained by just feeding data into a deep learning model. This is because the target signal is much weaker than the background signal and noise, resulting in a low signal-to-noise ratio. Therefore, artifacts removal is commonly adopted to remove artifacts so that the signal-to-noise ratio can be improved before feeding into a deep learning model. This is quite different compared to image or video processing, where image or video is directly fed into a deep learning model. To date, different kinds of deep learning models have been employed to process and classify EEG signal. Cecotti et al. used convolutional neural network (CNN) to extract features from steady-state visual evoked potential in 2008 \cite{cecotti2008convolutional}. Li et al. employed denoising autoencoder to classify two classes of motor imagery using EEG recorded from 14 electrodes on the sensorimotor cortex \cite{li2014deep}. Tsiouris et al. applied recurrent neural network (RNN) to capture sequential relationships for seizure detection \cite{Tsiouris2018}. A survey covering six EEG-based applications was done in 2019, where studies were reviewed separately for task type, model type and so on \cite{craik2019deep}. A more specialized survey on motor imagery classification can be found in \cite{al2021deep}. A distribution summary showing which disease is dominantly targeted in the studies of deep learning-based disease diagnosis can be found in \cite{rivera2021diagnosis}. If you want to read a survey on brain-computer interface (more beyond motor imagery), it can be found in Section 5 of \cite{gu2021eeg}. If a wide range of topics of deep learning in EEG is sought, this survey can be an option.

  Although EEG domain is far behind compared to the domains such as computer vision \cite{rawat2017deep} and speech recognition \cite{nassif2019speech} in terms of adopting deep learning, significant progress has been achieved in the last decade. It is time to summarize the achievements of deep learning in EEG for the past 10 years and discuss current existing issues and future directions. The searching criterion ["Deep Learning" AND "EEG" AND "Classification" OR "Recognition" OR "Identification"] was used for literature retrieval in the Web of Science in March 2020. After manual selection, 193 papers were included in this survey. During the revision in February 2021, we applied the same searching criterion to find newly-published literature after the previous searching and selected 20 papers to be included in this survey. After the acceptance, seven more papers were further included, but they were not used to update the figures and tables due to the constrained time. 
  
  As shown in Fig. \ref{fig_ratio_year}, the majority of these papers were published after 2017 while there was a rapid increase from the year of 2019. In 2019, the number of papers in the topic of brain-computer interface and disease detection are significantly more than the other topics. In 2020, the numbers of the published papers in more topics are rapidly increased, although disease detection is still a leading topic. The rapid increase of the published papers about deep learning in EEG is continued in 2021. The remainder of the survey is organized as follows. In Section II, artifacts removal is briefly introduced. This is followed by the detailed descriptions of all deep learning models which have been applied to EEG in Section III. In this section, we also mention the advantages and limitations of each deep learning model. Subsequently, the applications of deep learning in EEG are detailed along with publicly available EEG datasets used in these applications in Section IV. Finally, discussions are given and future directions are drawn at the end of the survey. All abbreviations used in this survey are listed in Table \ref{tab_abbr}.
	
\begin{figure}
\includegraphics[width=\textwidth]{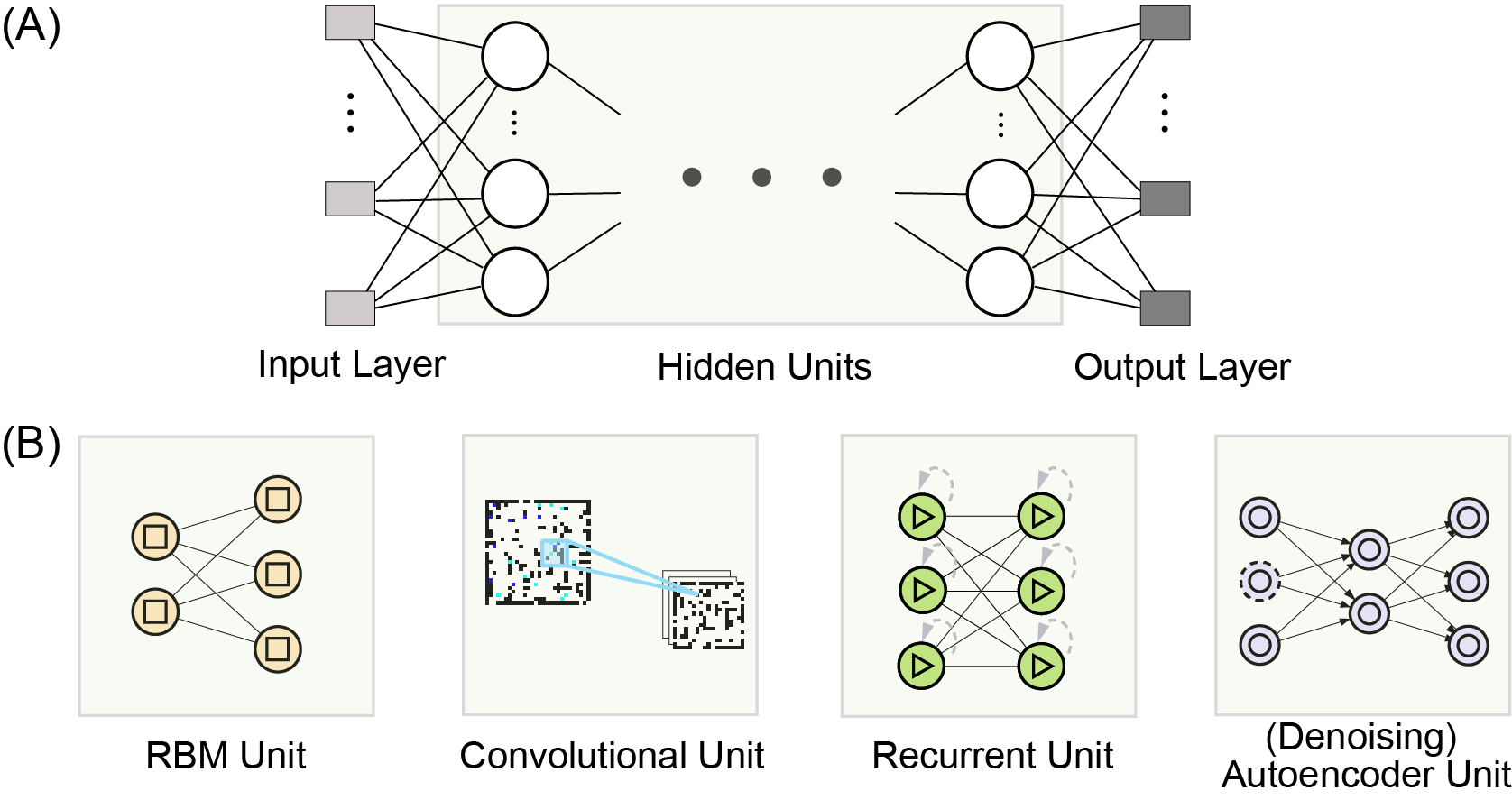}
\centering
		\caption{(A) Generic framework of a deep learning model. (B) Classical units that are employed in a deep learning model.}
\label{fig_overall_nets}
\end{figure}
	
	\section{Artifacts Removal}
	In general, artifacts are larger than that we intend to extract from EEG signal in terms of scale, leading to a low signal-to-noise ratio (SNR). In order to improve SNR, EEG signal is preprocessed to remove or mitigate the effect of artifacts on the signal before the signal is further processed. For example, a notch filter \cite{Li2017} is effective for eliminating the interference of power line. Independent component analysis \cite{Hosseini2017} is usually utilized to remove eye movements-related and muscular activity-related artifacts. Classical methods of artifacts removal and their targeted artifacts are summarized in Table \ref{tab_artifacts}. 
  
  When deep learning emerges, the step of artifacts removal is kept. EEG signal is preprocessed as usual to remove artifacts before inputting into a deep learning model. This is an effective way as all artifacts removal methods can be applied with deep learning models to be of both benefits inherited from the artifacts removal methods and deep learning models. This is also a natural and straightforward way that researchers are able to easily implement. However, an independent step of artifacts removal is not always necessary. The first several layers in a deep learning model could be functioned as artifacts removal, where noise is removed through the layers. To this end, a few attempts were done. For example, Supratak et al. inputted raw EEG data into a CNN for the classification of sleep stages. Their study showed that an acceptable performance can be achieved without an independent step of artifacts removal \cite{Supratak2017}. In addition, Bahador et al. mapped the correlation of EEG channels into a 2D space and used a CNN model to learn representations related to particular artifacts. With respect to artifact detection, this method outperformed spectrogram-based CNNs \cite{bahador_correlation-driven_2020}. Moreover, no auxiliary reference signal was required in their method.

	
	\section{Deep Learning Models}
	In this section, we describe each fundamental deep learning model. Their variants and combinations are not included as they share the similar rationale with fundamental models. A deep learning model is a hierarchical structure, comprising layers through which data are mapped into more and more abstract. Whatever a deep learning model is, there are an input layer, an output layer, and one or more hidden units (see Fig. \ref{fig_overall_nets}(A)). The hidden unit might be one of the layer structures illustrated in Fig. \ref{fig_overall_nets}(B) or their combinations. In the following subsections, we introduce classical deep learning models where typical units illustrated in Fig. \ref{fig_overall_nets}(B) are embedded.
	
	\subsection{Restricted Boltzmann Machine and Deep Belief Networks}
	A restricted Boltzmann machine (RBM) \cite{salakhutdinov2007restricted} is an undirected graph model (see Fig. \ref{fig_overall_nets}(B): RBM Unit), which has a visible layer $ \mathbf{v}=(v_{1},v_{2},\dots,v_{n}) $
	and a hidden layer  $ \mathbf{h}=(h_{1},h_{2},\dots,h_{n}) $.
	Connections exist only between visible layer $\mathbf{v}$ and hidden layer $\mathbf{h}$ and there are no connections between nodes within the visible layer or hidden layer. The energy function for an RBM is defined as: 
	
	\begin{equation}
	E(\mathbf{v}, \mathbf{h})=-\mathbf{v}^{T} \mathbf{W} \mathbf{h}-\mathbf{a}^{T} \mathbf{v}-\mathbf{b}^{T} \mathbf{h}
	\end{equation}
	where $\mathbf{W}$ is the weight matrix, $\mathbf{a}$ and $\mathbf{b}$ are bias vectors. The joint probability of $\mathbf{v}$ and $\mathbf{h}$ is constructed in terms of $E$:
	
	\begin{equation}
	P(\mathbf{v}, \mathbf{h})=\frac{1}{Z} e^{-E(\mathbf{v}, \mathbf{h})}
	\end{equation}
	where $Z$ is a normalizing constant defined as:
	\begin{equation}
	Z=\sum_{\mathbf{v}, \mathbf{h}} e^{-E\left(\mathbf{v}, \mathbf{h}\right)}
	\end{equation}
	The marginal distribution over the visible variables is obtained as:
	\begin{equation}
	P(\mathbf{v})=\frac{1}{Z} \sum_{\mathbf{h}} e^{-E(\mathbf{v}, \mathbf{h})} 
	\end{equation}
	The conditional probabilities can be described as:
	\begin{equation}
	P\left(h_{j}=1| \mathbf{v}\right)=\sigma\left(  \mathbf{W}_{j} \mathbf{v} + b_{j}\right)
	\end{equation}
	\begin{equation}
	P\left(v_{i}=1| \mathbf{h}\right)=\sigma\left(  \mathbf{W}_{i} \mathbf{h} + a_{i}\right)
	\end{equation}
	where $\sigma$ is logistic function defined as:
	\begin{equation}
	\sigma(x)=\left(1+e^{-x}\right)^{-1}    
	\end{equation}
	
	A deep belief network (DBN) is constructed by stacking multiple RBMs \cite{hinton2006fast}. Each RBM in the DBN is trained using an unsupervised manner at first. Then, the output of previous RBM is inputted into the next RBM. All RBMs are fine-tuned together by supervised optimization.
	
	\subsection{Convolutional Neural Network}
	Convolutional neural network (CNN) \cite{krizhevsky2012imagenet} is good at capturing spatial information of data (see  Fig. \ref{fig_overall_nets}(B): Convolutional Unit). Most CNNs consist of two types of layers: convolutional layer, pooling layer.

	In specific, a convolutional layer has filters $k_{ij}^{l}$, the size of which is usually much smaller than the dimension of input data and forms a locally connected structure. Filter at layer $l$ can produce feature maps $ \mathbf{X}_{j}^{l} $ by convolving with the input $\mathbf{X}_{i}^{l-1}$ plus biases $b_{j}^{l}$. These features are subjected to a non-linear transformation $ f(\cdot) $ and can be mathematically expressed as:
	\begin{equation}
	\mathbf{X}_{j}^{l}=f\left(\sum_{i=1}^{M^{l-1}}\mathbf{X}_{i}^{l-1} * k_{ij}^{l}+b_{j}^{l}\right)
	\end{equation}
	Where $M^{l-1} $ represents the number of feature maps in layer $ l-1$, and $*$ denotes convolution operation.
	
	A pooling layer is responsible for feature selection and information filtering. Two kinds of pooling operations are widely used: max pooling and average pooling. In max pooling, maximum value is mapped from a sub-region by pooling operator. In average pooling, the average value of a sub-region is selected as the result. A fully-connected layer is usually added at the last part of a CNN in the case of classification. It transforms a long 1D vector and outputs to the next layer (usually softmax).

	Weight sharing and sparse connections are two basic strategies in CNN models, which lead to dramatic reduction in the number of parameters. These strategies are helpful to reduce training time and enhance training effectiveness. Moreover, they also mitigate the overfitting problem while retaining a good capability of complex feature extraction. 
	
	\subsection{Recurrent Neural Networks}
	Recurrent neural network (RNN) \cite{lipton2015critical} was developed to deal with sequential data because of its unique recurrent structure (see  Fig. \ref{fig_overall_nets}(B): Recurrent Unit), which allows previous outputs to be used as inputs while having hidden states. It is widely used in applications that need to extract sequential information, such as natural language processing, speech recognition, and EEG classification.
	
	\subsubsection{GRU}
	Gated Recurrent Unit (GRU) \cite{chung2014empirical} has two gates, reset $\mathbf{r}_{t}$ and update $\mathbf{z}_{t}$. Let $\mathbf{x}_{t}$ be the input at time step $ t $ to a GRU layer and $\mathbf{h}_{t}$ be the output vector. The output activation is a linear interpolation between the activation from the previous time step and a candidate activation $\hat{\mathbf{h}}_{t}$.
	\begin{equation}
	{\mathbf{h}_{t}=\mathbf{z}_{t} \odot \mathbf{h}_{t-1}+\left(1-\mathbf{z}_{t}\right) \odot \tilde{\mathbf{h}}_{t}}
	\end{equation}
	where $\mathbf{z}_{t}$ decides the interpolation weight, which is computed by:
	\begin{equation}
	{\mathbf{z}_{t}=f\left(\mathbf{W}_{z} \mathbf{x}_{t}+\mathbf{U}_{z} \mathbf{h}_{t-1}+\mathbf{b}_{z}\right)} 
	\end{equation}
	where $\mathbf{W}$ and $\mathbf{U}$ are weight matrices for the update gate, $\mathbf{b}$ is a bias vector, and $f(\cdot)$ is a non-linear function (usually sigmoid function). The candidate activation is also controlled by an additional reset gate and computed as follows:
	\begin{equation}
	{\tilde{\mathbf{h}}_{t}=g \left(\mathbf{W}_{h} \mathbf{x}_{t}+\mathbf{U}_{h}\left(\mathbf{r}_{t} \odot \mathbf{h}_{t-1}\right) + \mathbf{b}_{h}\right)}
	\end{equation}
	where $\odot$ represents an element-wise multiplication and $g(\cdot)$ is often a non-linear tanh function. The reset gate is computed in a similar manner as the update gate:
	\begin{equation}
	{\mathbf{r}_{t}=f\left(\mathbf{W}_{r} \mathbf{x}_{t}+\mathbf{U}_{r} \mathbf{h}_{t-1}+\mathbf{b}_{r}\right)}
	\end{equation}
	
	\subsubsection{LSTM}
	
	Different from GRU, Long Short-Term Memory (LSTM) \cite{hochreiter1997long} has three gates, input $\mathbf{i}_{t}$, output $\mathbf{o}_{t}$, and forget gates $\mathbf{f}_{t}$.
	Each LSTM cell has an additional memory component $\mathbf{c}_{t}$. 
	The gates are calculated in a similar manner as the GRU but LSTM has additional memory components.
	
	\begin{equation}
	\mathbf{i}_{t}  =f\left(\mathbf{W}_{i} \mathbf{x}_{t}+\mathbf{U}_{i} \mathbf{h}_{t-1}+\mathbf{b}_{i}\right)
	\end{equation}
	
	\begin{equation}
	\mathbf{o}_{t}  =f\left(\mathbf{W}_{o} \mathbf{x}_{t}+\mathbf{U}_{o} \mathbf{h}_{t-1}+\mathbf{b}_{o}\right)
	\end{equation}
	
	\begin{equation}
	\mathbf{f}_{t}  =f\left(\mathbf{W}_{f} \mathbf{x}_{t}+\mathbf{U}_{f} \mathbf{h}_{t-1}+\mathbf{b}_{f}\right) 
	\end{equation}
	A memory component is updated by forgetting the existing content and adding a new memory component as:
	\begin{equation}
	\mathbf{c}_{t}=\mathbf{f}_{t}  \odot\mathbf{c}_{t-1}+\mathbf{i}_{t} \odot \hat{\mathbf{c}}_{t}
	\end{equation}
	where $\hat{\mathbf{c}}_{t}$ can be computed by:
	\begin{equation}
	\hat{\mathbf{c}}_{t}=g\left(\mathbf{W}_{c} \mathbf{x}_{t}+\mathbf{U}_{c} \mathbf{h}_{t-1} +\mathbf{b}_{c}\right)
	\end{equation}
	The updated equation for the memory component is controlled by the forget and input gates. Then, the output of the LSTM unit is computed from the memory modulated by the output gate according to the following equation:
	\begin{equation}
	\mathbf{h}_{t}=\mathbf{o}_{t}  \odot g\left(\mathbf{c}_{t}\right)
	\end{equation}
	
	\subsection{Autoencoder and Stacked Autoencoder}
	 Autoencoder (AE) is a symmetrical structure with two layers \cite{rumelhart1985learning} (see Fig. \ref{fig_overall_nets}(B): Autoencoder Unit). 
	 
	 An encoder learns latent representation from the input data while a decoder restores the latent representation as close to the input data as possible.  The goal of an autoencoder is to minimize the reconstruction error between the input and the output. 
	
	Given the inputs $\mathbf{x} \in \mathbf{R}$, the encoding process first maps it into a latent representation $\mathbf{h} \in \mathbf{R}$ through a weight matrix $\mathbf{W}_{v}$, bias $\mathbf{b}_{v}$, and an activation function $f(\cdot)$:
	\begin{equation}
	\mathbf{h}=f\left(\mathbf{W}_{v} \mathbf{x}+\mathbf{b}_{v}\right)
	\end{equation}
	Then the decoding process transforms the latent representation $\mathbf{h}$ into the reconstruction $\mathbf{y}$ through a weight matrix $\mathbf{W}_{h}$, bias $\mathbf{b}_{h}$, and an activation function $g(\cdot)$:
	\begin{equation}
	\mathbf{y}=g\left(\mathbf{W}_{h} \mathbf{h}+\mathbf{b}_{h}\right)
	\end{equation}
	To simplify the network architecture, the tied weights strategy $\mathbf{W}_{v}=\mathbf{W}_{h}=\mathbf{W}$ are usually employed. The parameters to be determined are $\{\mathbf{W},\mathbf{b}_{v}, \mathbf{b}_{h}\}$. The training of an autoencoder is to minimize the loss:
	\begin{equation}
	\arg \min _{\mathbf{W}, \mathbf{b}_{v}, \mathbf{b}_{h}} \mathcal{J}\left(\mathbf{W}, \mathbf{b}_{v}, \mathbf{b}_{h}\right)
	\end{equation}
	Given the training samples $\mathbf{D}_{n}$, the loss function is defined as:
	\begin{equation}
	\mathcal{J}\left(\mathbf{W}, \mathbf{b}_{v},\mathbf{b}_{h}\right)=\frac{1}{N_{D_n}}\sum_{\mathbf{x}\in \mathbf{D}_{n}}
	L(\mathbf{x}, \mathbf{y})
	\end{equation}
	where $L$ is the error of the reconstruction and $N_{D_n}$ is the number of the training samples.
	
	Stacked autoencoder (SAE) is a neural network, where autoencoders are connected one another to form a cascade.

	\subsection{Others}
	In addition to the aforementioned models, there are other models aiming to solve particular shortcomings existing in the above models. For example, capsule network (CapsNet) was proposed to overcome the shortcoming that CNN does not well capture the relationships between the parts of an image \cite{sabour2017dynamic}. When it applied to fMRI \cite{wang2020multikernel} and EEG \cite{Chao2019}, it is expected to capture comprehensive relationships among brain regions, channels, or frequencies, and so on. To shorten training time, extreme learning machine (ELM) was proposed, where the weights of hidden layers are randomly assigned and fixed during the training \cite{Ding2015}. Weight randomization is also implemented in echo state network (ESN) \cite{Bozhkov2017}. ESN is a recurrent neural network where the weights of hidden layers are randomly and sparsely assigned and fixed while the weights of output layer can be tuned. Spiking neural network (SNN) is a biologically inspired model and has been used to explore brain activity patterns in \cite{Doborjeh2016}. Deep polynomial network (DPN) uses a quadratic function to process its inputs and is able to learn features between different samples or dimensions. It was implemented in \cite{Lei2019} to utilize features from multiple views for motor imagery classification, including common spatial pattern, power spectral density, and wavelet packet transform. In addition, some variants of deep learning models were proposed by using different training strategies, such as generative adversarial network.    
	
	\section{Applications}
\begin{figure}[!hbt]
	\centering
		\includegraphics[width=0.9\textwidth]{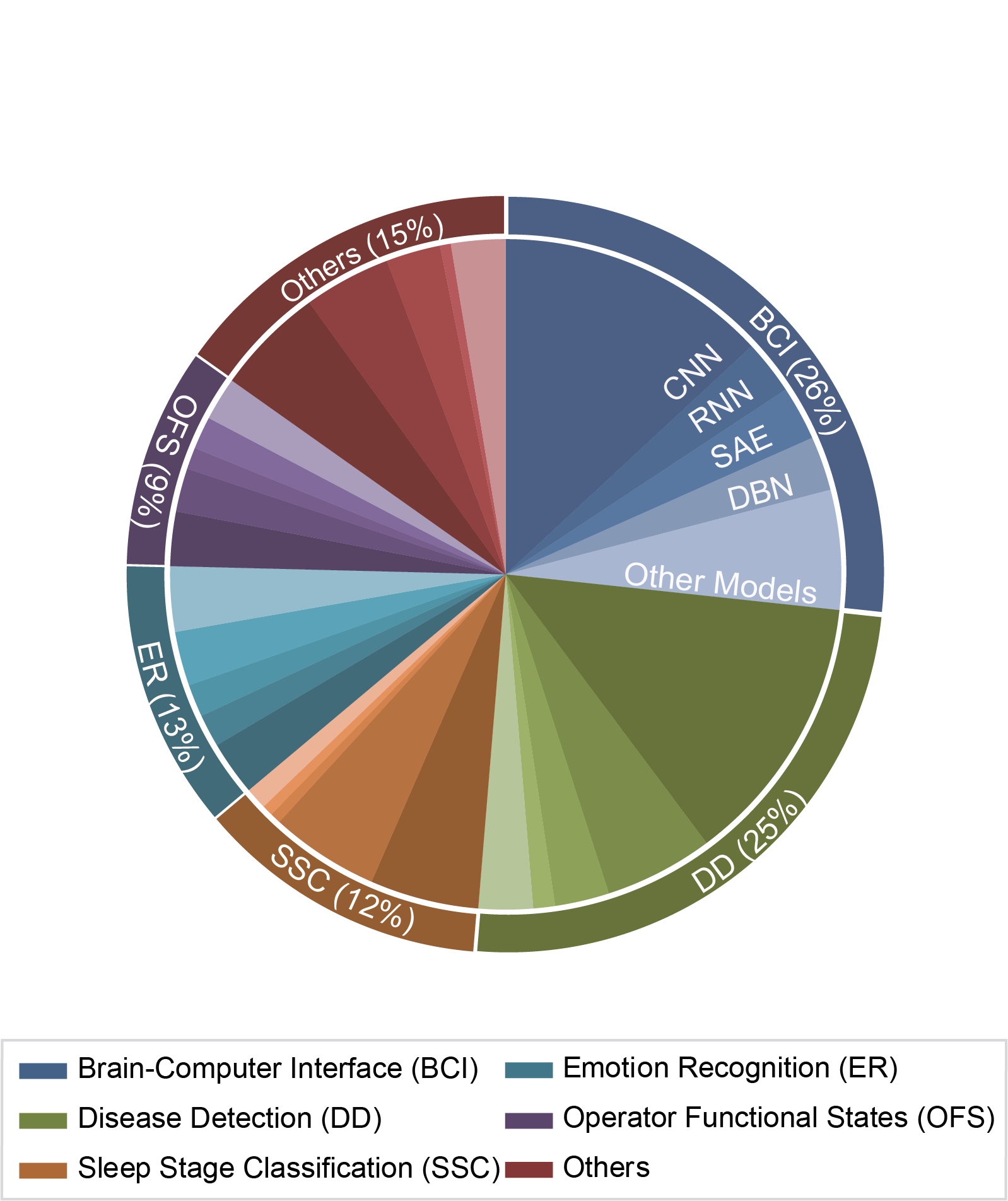}
		\caption{Percentages of application topics and deep learning models. The outer ring represents paper percentages for each topic. The models within each topic are distinguished from the darkest to lightest colors, which stand for CNN, RNN, SAE, DBN, and other models in order.}
		\label{fig_ratio_overall}
	\end{figure}

	We summarized applications, in which deep learning was utilized for EEG processing and classification, in this section. For your convenience, we group diverse applications into six topics, which are brain-computer interface (see Table \ref{tab_bci} for the details of studies), disease detection (see Table \ref{tab_disease}), emotion recognition (see Table \ref{tab_emotion}), operator functional states (see Table \ref{tab_ofs}), sleep stage classification (see Table \ref{tab_slepp}), as well as the applications other than above topics (see Table \ref{tab_other_applications}). According to statistics, the majority of selected papers belong to the topics of brain-computer interface (account for 26\%) and disease detection (account for 25\%). The percentages of each topic and the percentages of each model used in each topic are illustrated in Fig. \ref{fig_ratio_overall}. In addition, we collected the information of the publicly available datasets which had been used in the studies and listed them in Table \ref{tab_datasets}.
	
	\subsection{Brain-Computer Interface}
\begin{figure}[!hbt]
	\centering
		\includegraphics[width=\textwidth]{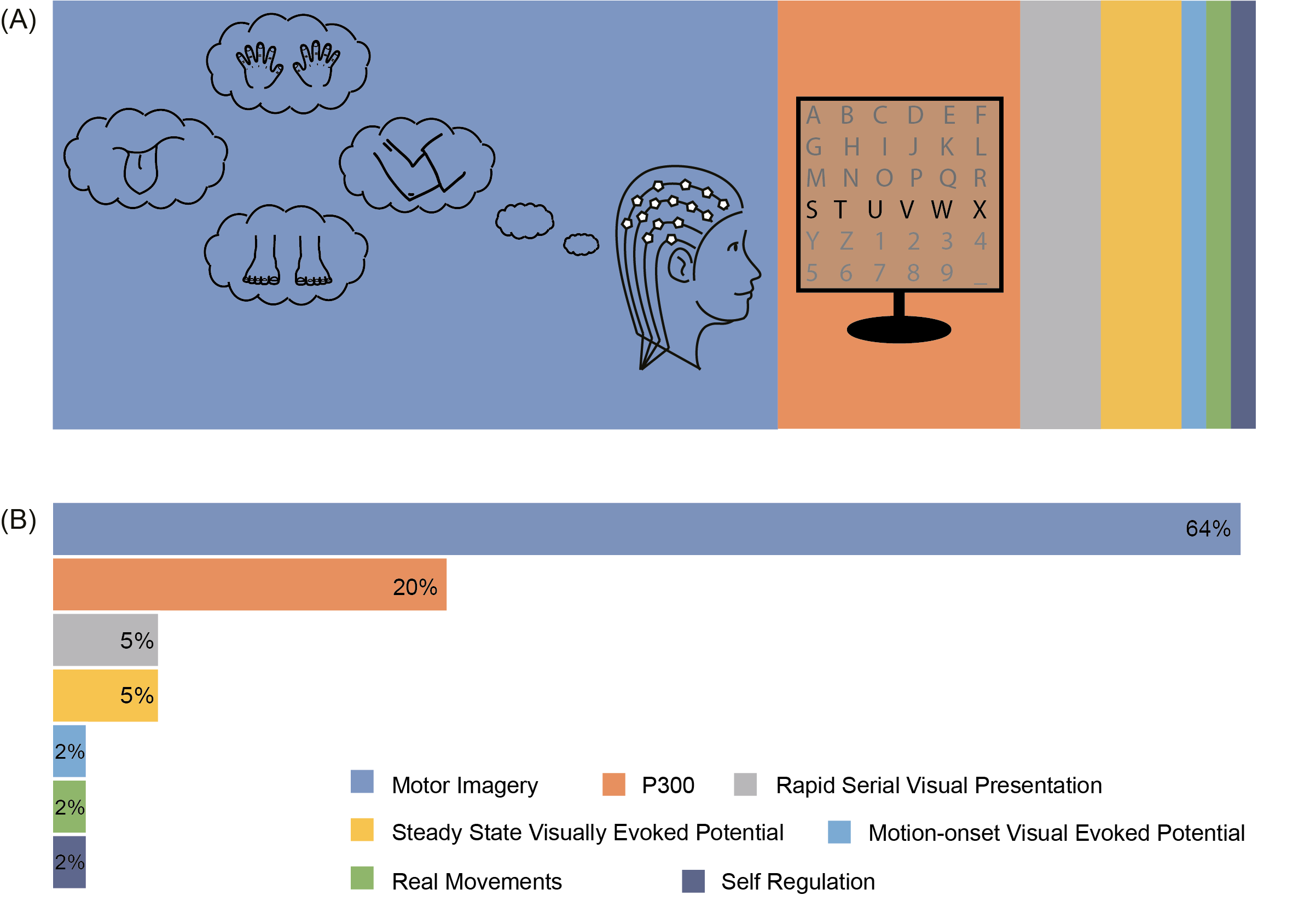}
		\caption{(A) Paradigms of brain-computer interface. (B) Percentages of the selected papers for each paradigm by the year of 2020}
		\label{fig_bci_branches}
	\end{figure}

	A brain-computer interface (BCI) can be defined as a system that decodes brain activity and translate user's intentions into messages or commands for the purposes of communication or the control of external devices, and more. In this topic, deep learning was mainly applied to establish motor imagery (MI)- and P300-based BCIs (see Fig. \ref{fig_bci_branches}). 
	
	Transfer learning is utilized to mitigate the cost of re-training or solve the problem of data lack in the target domain. A deep learning model trained on the data collected from a session or a subject can be transferred to classify/recognise the data of another session or another subject with a fine-tuning. In some cases, the fine-tuning is omitted. In general, the fine-tuning positively contributes to the performance. The extent of fine-tuning was investigated in a recent study\cite{zhang2021adaptive}. It shows that the best performance of motor imagery classification was achieved when all layers were tuned except the first hidden layer under the condition of a low learning rate. Another study comparing cross-session transferring and cross-subject transferring demonstrated that the cross-session transferring was feasible and the cross-subject transferring was inefficient \cite{Lu2017}. With the combination of transfer learning and CNN, Hang et al. proposed a deep domain adaption network \cite{Hang2019}. They used maximum mean discrepancy to minimize the distribution discrepancy between target and source subjects and used the center-based discriminative feature learning method to make deep features closer to corresponding class centers. The evaluation on BCI Competition datasets (i.e., Dataset IVa of Competition III and Dataset IIa of Competition IV) demonstrated a good classification performance. In the study of cross-subject transferring \cite{Zhao2019}, network weights were transferred. Dose et al. used a pool of data to obtain a universal model of CNN \cite{Dose2018}. This model was then adapted based on a small amount of data from a subject before applying to this subject. Their results showed that an average improvement of 6$\sim$9\%  was achieved for motor imagery classification in terms of classification accuracy.

	Transferring can also be conducted between domains. A CNN-based model (VGG-16) trained on image data (the data from ImageNet) was transferred to recognize EEG data by freezing the parameters in the first several layers and fine-tuning the parameters in the last several layers using an EEG dataset \cite{Xu2019a}. The performance was better than that of support vector machine. Similar to the domain of image recognition, the amount of EEG data can also be increased by augmentation procedure. Li et al. produced new samples by adding noise into EEG data \cite{Li2019c}. They claimed that adding noise into amplitudes of power spectra was superior to that adding noise into EEG time series in terms of classification accuracy. Zhang et al. used intrinsic mode functions derived from empirical mode decomposition to generate new EEG samples so that the total number of samples was increased \cite{Zhang2019a}.

	Classical models such as CNN and RNN were originally developed for image or speech recognition, so they did not well match the characteristics of EEG signal. They should be adapted before applying to EEG recognition. Li et al. designed a CNN-based network consisted of three blocks to capture spatial and temporal dependencies \cite{Li2019c}. Multi-channel raw EEG signals were fed into temporal convolutional layer and spatial convolutional layer successively in the first block. In the second block, a standard convolutional layer and a dilated convolutional layer were utilized to extract temporal information at different scales while reducing the number of parameters. The extracted features were finally used for motor imagery classification in the third block. In another CNN-based network \cite{Li2019a}, a layer was fed by all outputs from previous layers and its output was inputted to all following layers. By using such dense inter-layer connections, information loss could be reduced. In \cite{Zhang2019a}, EEG signals were transformed into tensors and fed into a CNN-like network where convolution were replaced with complex Morlet wavelets, resulting in parameter reduction. Wavelet kernel was also used to learn time-frequency features \cite{Zhao2019}. Their results demonstrated that wavelet kernels can provide faster convergence rate and higher classification accuracy compared to plain CNN. Alazrai et al. used CNN to extract features from time-frequency images, which were transformed using a quadratic time-frequency distribution \cite{Alazrai2019}. The methods were compared to a support vector machine, and it suggested that CNN can achieve good performance in MI tasks of the same hand.

	In order to accelerate the training course and alleviate the overfitting problem, Liu et al. adjusted the number and position of batch normalization layers in a CNN-based network for P300 detection \cite{Liu2018}. Kshirsagar et al. employed leaky rectified linear unit activation function at each convolutional layer \cite{Kshirsagar2019}. To evaluate whether the number of convolutional layers needs to be adjusted for different BCI tasks and find out an optimal structure, Lawhern et al. compared networks with different numbers of convolutional layers \cite{Lawhern2018}. Their results showed that deep CNN (i.e., five convolutional layers) tended to perform better on the oscillatory BCI dataset than on the event-related potential BCI dataset, while shallow CNN (i.e., two convolutional layers) achieved better performance on the event-related potential BCI dataset. Apart from CNN, Lu et al. used a DBN (i.e., three RBMs and an output layer) to extract features of motor imagery \cite{Lu2017}. Some studies aimed to compare performances of different deep learning models. For example, Pei et al. compared SAE and CNN in the classification of reaching movements \cite{Pei2018}. They found that SAE was better than CNN and suggested that poorer performance in CNN might be due to the lack of training data. One year later, another study comparing between these two models showed that SAE had satisfactory performance in some trials, but inefficient to those trials of the subjects who were less attentive in P300 detection, while CNN performed well in terms of accuracy and information transfer rate \cite{Kshirsagar2019}.

	The combination of deep learning model and traditional model or the mixture of two or more types of deep learning models is applied to EEG classification. For example, SAE was combined with support vector machine to classify EEG signal \cite{Kundu2019}. SAE was also combined with CNN to develop a new model \cite{Tabar2017}, where CNN layers were used to extract features from 2D time-frequency images (obtained by Fourier transform over EEG signals) and SAE was further used to extract features. In \cite{Amin2019}, the features extracted by CNN were fed into an autoencoder for cross-subject MI classification. This combination achieved a better accuracy for the cross-subject classification, but worse for the subject-specific classification, compared to the combination of CNN and multilayer perceptron (MLP). Zhang et al. presented a hybrid network comprised of CNN and LSTM, in which EEG signals were sequentially processed through common spatial pattern, CNN, and LSTM \cite{Zhang2019f}. The idea of using CNN and LSTM to extract spatial and temporal features was also conceived by Yang et al. \cite{Yang2018a}. However, they inserted a discrete wavelet transformation (DWT) between CNN and LSTM, which led to better performance in the MI classification compared to that of pure combination of CNN and LSTM.

	In addition to P300- and MI-based BCIs, deep learning models also applies to the other BCIs, including motion-onset visual evoked potentials \cite{Ma2017} and self-paced reaching movements \cite{Pei2018}. Nguyen et al. developed a steady state visually evoked potential (SSVEP)-based BCI speller system, in which only one channel was used \cite{Nguyen2019}. They used fast Fourier transform to extract features from this channel and then fed the features into a CNN model. According to their results, frequency resolution and time window length influence classification performance. The frequency resolution of 0.0625 Hz and time window of 2s were optimal for the five-class classification \cite{Nguyen2019}. Waytowich et al. proposed a compact CNN to deal with asynchronous problem in SSVEP classification \cite{Waytowich2018}. It outperformed canonical correlation analysis (CCA) and combined-CCA.

	\begin{figure}
		\centering
		\includegraphics[width=\textwidth]{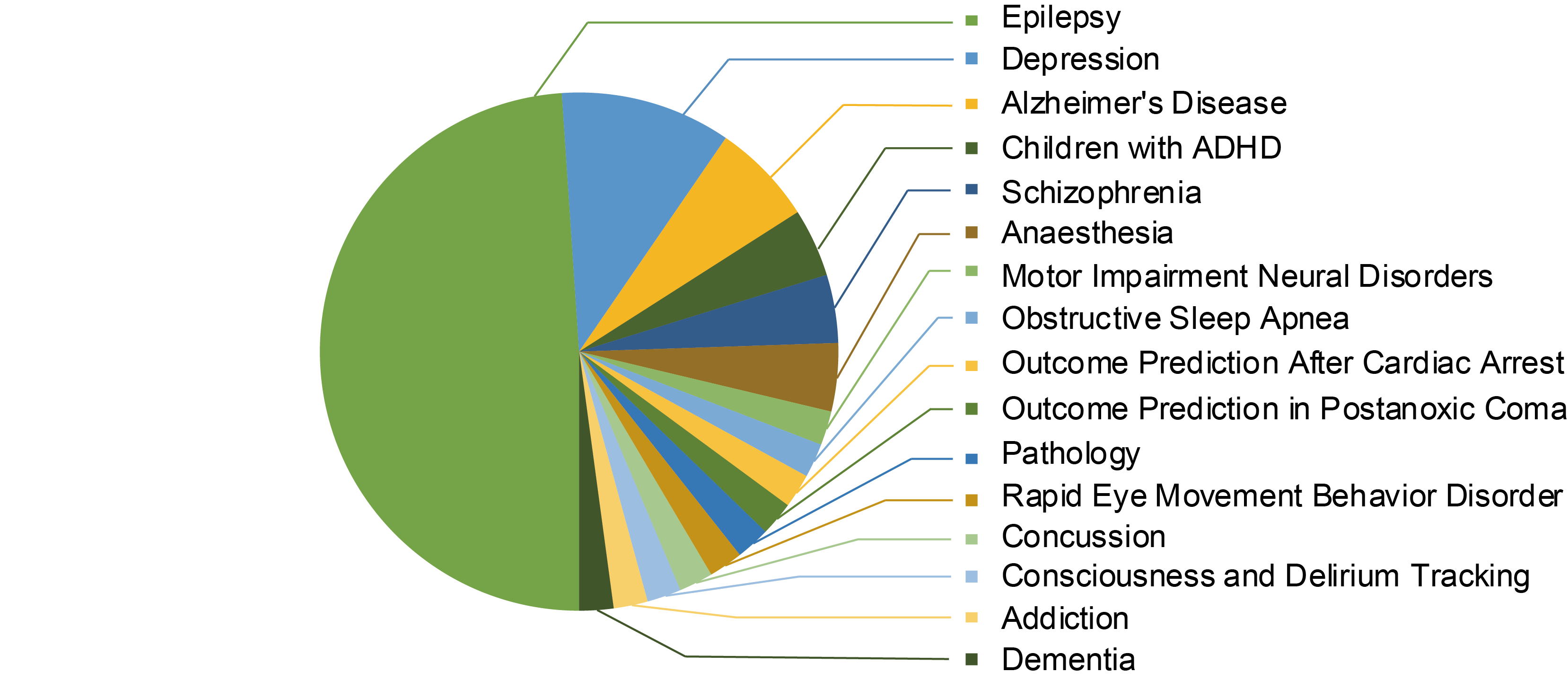}
		\caption{Percentages of the selected papers across diseases.}
		\label{fig_disease_branches}
	\end{figure}

	\subsection{Disease Detection}
	Machine learning could benefit disease diagnosis by providing assistant information and preliminary diagnostic results. In this topic, deep learning models were also widely employed to detect a variety of diseases (see the distribution of the selected papers over diseases in Fig. \ref{fig_disease_branches}). In this subsection, commonly used models and model designing strategies were introduced at first, including the examples of single or hybrid models, as well as the detailed architecture (e.g., layer settings). Afterwards, we described other techniques that have an influence on the performance of deep learning.

  CNN is a deep learning model, which has been widely adopted for the detection of brain diseases (e.g., seizure detection \cite{Acharya2018a} and schizophrenia identification \cite{Oh2019}). Cao et al. stacked multiple CNNs to classify epileptic signals. In this study, the proposed model was compared to a few classification algorithms (i.e., Support Vector Machine (SVM), k-Nearest Neighbours (kNN), ELM) under different conditions (i.e., 1. Two-class, seizure/non-seizure; 2. Three-class, interictal/preictal/ictal; 3. Five-class, interictal/three preictal states/ictal) \cite{cao_epileptic_2020}. To enhance the performance of epilepsy classification, original binary labels, namely interictal epileptiform discharge (IED) and non-IED, were converted into multiple labels used for model training \cite{Antoniades2017}. Specifically, samples were further divided into five subclasses according to spatial distribution and morphology of EEG waveforms and were then fed into a CNN model for the training. A new sample was first classified to one of these subclasses and then the final classification result (IED versus non-IED) was obtained by applying a threshold at the last layer.   Compared to the CNN model training with binary labels, the training with further finer tags could enhance the discriminative power of the model and led to better performance in the most subjects.

  When CNN is combined with other models, classification performance can be improved. In \cite{Wen2018}, CNN and autoencoder (AE) were combined to learn robust features in an unsupervised way. The integrated network had an encoder consisting of convolution and down-sampling and a decoder consisting of deconvolution and up-sampling. Their results demonstrated that CNN+AE is superior to principal component analysis (PCA) and sparse random projection (SRP) in epilepsy related feature extraction. In \cite{Daoud2019}, a hybrid model combining CNN, AE, and LSTM achieved remarkable prediction of seizure. Combined deep learning model was used for pre-training and latent representation learning. By this, the accuracy of focal and non-focal classification was improved \cite{Daoud_deep_2020}. However, model combination is not always positive to the performance improvement. Some studies showed that performance may decline in some cases. For instance, Mumtaz et al. combined CNN and LSTM to detect unipolar depression. Their results showed that the hybrid model did not outperform single model of CNN \cite{Mumtaz2019}.

  Beyond the selection of deep learning models, model settings also vary across studies. Tsiouris et al. found that overfitting problem can be mitigated by shuffling input EEG segments, which could replace the dropout role partially \cite{Tsiouris2018}. Qiu et al. applied data corruption in the stacked autoencoder for seizure detection \cite{Qiu2018}. Specifically, they designed a denoising sparse autoencoder, in which some of the input data were set to zero. This improved model robustness and reduced overfitting problem. In addition, performance is also influenced by the condition of data recording. Mumtaz et al. found that unipolar depression can be more accurately detected using the EEG recorded under the condition of eyes open compared to that of eyes closed \cite{Mumtaz2019}. In the study of attention deficit hyperactivity disorder (ADHD) detection using a CNN model, EEG signals at different channels were rearranged to make adjacent channels together in the connectivity matrix to improve accuracy \cite{Chen2019e}. Moreover, Tsiouris et al. shuffled interictal and preictal segments of EEG to avoid the overfitting in seizure detection \cite{Tsiouris2018}. Yuan et al. used a channel-aware module to enhance the capability of feature learning and concentrate on important and relevant EEG channels \cite{Yuan2019a}. Daoud et al. computed the statistical variance and entropy of the channels, and selected those with the highest variance entropy product for seizure prediction \cite{Daoud2019}.

  The performance of deep learning for disease detection is affected by EEG data arrangement. For example, EEG data are reshaped into 2D format before inputting into a deep learning model. In \cite{Ieracitano2019}, EEG data were transformed into 2D images of spectral powers. Then, these images were fed into a CNN network for distinguishing Alzheimer’s disease and mild cognitive impairment from healthy controls. To differentiate patients with schizophrenia \cite{Naira2019}, Pearson correlation coefficients were calculated between channels and assembled as a correlation matrix. Correlation matrices of each subject were fed into a CNN network. Moreover, fast Fourier transform \cite{Vrbancic2018} and continuous wavelet transform \cite{Turk2019} were used to transform EEG data into 2D images for motor impairment neural disorders and epilepsy classification, respectively. Wei et al. further converted 2D images into 3D stacked images according to the mutual correlation intensity between channels \cite{Wei2018}. To utilize comprehensive information from different data forms, Tian et al. used three CNNs to respectively obtain features existing in the time, frequency, and time-frequency domain, and then ultilized these features for seizure detection \cite{Tian2019}. By comparing with the methods that ultilizing features from only one domain, the proposed method exhibited better performance. According to the study comparing among raw EEG signal, Fourier transform, wavelet transform, and empirical mode decomposition, raw signals and empirical mode decomposition were better than the others in distinguishing focal EEG from non-focal EEG, while Fourier transform was best in ictal and non-ictal classification \cite{San-Segundo2019}. To handle the problem of inadequate data, sliding time window was used to split continuous EEG signal into segments with partial overlapping to increase the data amount in \cite{Ullah2018}. Cao et al. developed an interactive system to help experts label the new data, and the data can be added to fine-tune the deep learning model to gradually improve the interictal-ictal continuum classification accuracy \cite{Cao2019}. 
  
	\begin{figure}[!htb]
	\centering
		\includegraphics[width=\textwidth]{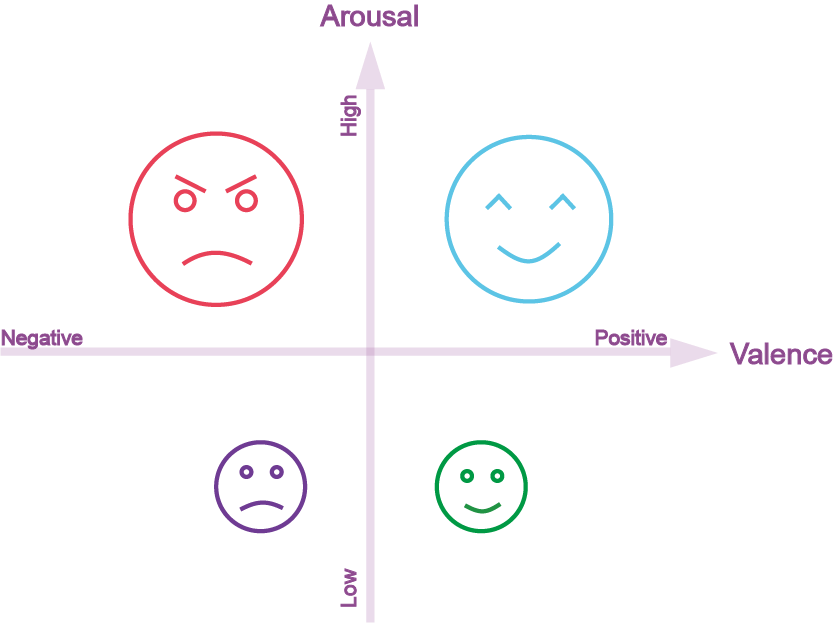}
		\caption{Four illustrative emotions classified based on the scores of arousal and valence.}
		\label{fig_emotion_distribution}	
	\end{figure}

	\subsection{Emotion Recognition}
	Emotion conveys lots of underlying information during conversations and is part of communication between people. People can understand emotion by reading facial expression, voice tone, and gestures. From the perspective of artificial intelligence, emotion can be recognized based on the data of facial expression \cite{Zhang2019b}, eye movement measures \cite{Zheng2019}, EEG \cite{Al-Nafjan2017}, or galvanic skin response signal \cite{Kwon2018}. According to the arousal and valence, emotion can be categorized into different classes (see Fig. \ref{fig_emotion_distribution}). Based on the statistics of the included papers in this survey, the studies mainly aimed to classify three classes (i.e., positive, neutral, and negative) or more classes (partitioned based on the scores of arousal and valence). Within these papers, the datasets named 'SEED' \cite{zheng2015investigating} and 'DEAP' \cite{koelstra2011deap} were frequently used to evaluate deep learning models for emotion recognition.
  
  SEED dataset was published by the BCMI laboratory at the Shanghai Jiao Tong University \cite{zheng2015investigating}. For this dataset, 62 channels were used to collect EEG data from 15 subjects when they were watching positive, negative, and neutral video clips. The data were collected from the subjects three times with an interval of one week or longer. Thus, it enables cross-session investigations. Zheng et al. demonstrated the stable patterns of EEG signals over time for emotion recognition \cite{zheng_identifying_2017}. Besides, they found that differential entropy could provide better performance than other features such as differential asymmetry and rational asymmetry. Using this dataset, Yang et al. proposed a hierarchical network which consists of subnetwork node, and this method boosted 5\%-10\% accuracy \cite{ yang_eeg-based_2018}. Li et al. trained a CNN and accomplished around 88\% of recognition accuracy based on features of the gamma band \cite{Li2018}. Zhang et al. proposed a two-layer RNN model to extract spatial and temporal features, respectively. The first layer of their model is an RNN layer that takes EEG signals from electrodes as inputs. The outputs of the first layer were concatenated along the time dimension and fed into the second RNN layer. The performance evaluated on the SEED dataset was 89.5\% \cite{Zhang2019b}. In \cite{Zeng2019}, Zeng et al. used an architecture that adapted from SincNet (a CNN-based network proposed for speaker recognition \cite{ravanelli2018speaker}) to classify emotion. Their results demonstrated that the adapted SincNet (i.e., three convolutional layers and three fully connected layers) was promising for emotion classification, reaching an accuracy of around 95\% as evaluated on the SEED dataset. 
  
  Another dataset named 'DEAP' \cite{koelstra2011deap}, was collected from 32 subjects when they watched 40 one-minute-long music videos. Perceptual emotion was assessed in terms of arousal, valence, liking, and dominance. Studies using this dataset have showed that deep learning was successful and effective to classify emotion categories based on EEG. \cite{Al-Nafjan2017}, \cite{Choi2018}. Even using raw EEG as the input, LSTM achieved an acceptable accuracy of around 85\% in the emotion classification \cite{Alhagry2017}. In \cite{Bagherzadeh2018}, various handcrafted EEG features (e.g. sample entropy, mean, and power spectral density) were fed into three stacked autoencoders in a parallel way for voting. Chao et al. also designed a parallel architecture to process EEG signal. However, they used DBN as the basic unit \cite{Chao2018}. To improve the classification performance and utilize strengths of different models. Li et al. combined CNN and LSTM to extract representations from multi-channel EEG, in which CNN was used to learn inter-channel and inter-frequency correlation while LSTM was used to extract contextual information \cite{Li2017a}. The model combination was also used in \cite{yin2021eeg}, where feature extraction was done by graph convolutional networks, temporal information was memorized by LSTM, and classification was done by a SVM. The same idea of model combination was also used in \cite{topic2021emotion}, where CNN was used for feature extraction.

Besides the two commonly used datasets (i.e., SEED and DEAP), Serap Aydın used affective video clips to induce nine emotional states (fear, anger, happiness, sadness, amusement, surprise, excitement, calmness, and disgust) and investigated gender effect on emotion recognition \cite{aydin_deep_2020}. This paper revealed that emotion is more affected by individual experience than gender. Zhu et al. designed an experiment to explored the emotion in the scenario of two-person interaction. In their experiment, two person need to rate their emotions induced by the same piciture one by one. They extracted the intra-brain and inter-brain phase synchronization features from emotional EEG signals and applied a CNN model to evaluate \cite{zhu_eeg-based_2020}. As we know, deep learning needs parameter tuning and it is time-consuming. To mitigate this problem, various strategies were proposed. Hemantha et al. modified the back-propagation neural network by arranging layers in a circular manner that the output can access the parameters of the input and hidden layers \cite{Hemanth2018}. This modification reduced convergence time by around 20\%. Jirayucharoensak et al. used principal component analysis for dimension reduction to lower computation cost \cite{Jirayucharoensak2014}. Gao et al. utilized gradient priority particle swarm optimization to optimize parameters of a CNN model \cite{Gao2020}.

	\subsection{Operator Functional States}
	The operator functional states (OFS) describe the mental states of operators in specific working conditions \cite{Yin2018}. Two of them are mental workload and mental fatigue. In specific, mental workload is a measure of cognitive resources consumed in the human working memory while mental fatigue is identified by an accumulated process of a disinclination of effort and drowsiness. To date, deep learning was used to identify mental states based on EEG signal. For example, drivers' \cite{Ma2019} \cite{Zeng2018} \cite{Gao2019} \cite{Chai2017} and pilots' \cite{Wu2019a} fatigue was monitored for the purposes of preventing fatigued operation.

	Generalization is one of the important metrics to evaluate a model. In the classification of operator functional states, large variance across subjects is challenging. Many studies employed subject-specific classifiers. For example, Tao et al. fused multiple ELMs and Naive Bayesian model to build a subject-specific classifier. This ensemble model with fine-tuned hyper-parameters was of the higher subject-specific accuracy in mental workload assessment \cite{Tao2019}. In the study of \cite{Zhang2017}, Zhang et al. selected the most relevant EEG channels for each subject and used these subject-specific channels for calculating weights between the input layer and the first hidden layer in the DBN. In contrast to the subject-specific models, the cross-subject model aims to have a general model for tolerating variance of subjects. For example, Heron et al. used multi-path convolutional layers and bi-directional LSTM layers to learn frequency and temporal features over subjects. This model achieved low variance in performance across subjects and showed better generalization compared to subject-specific models \cite{Hefron2018}. Another cross-subject model was proposed using an adaptive DBN with the weights of the first hidden layer iteratively updated to track the EEG changes in a new subject \cite{Yin2017b}. When different tasks were used to induce mental workload, the induced workload might be variable across tasks. The cross-task workload classification was made by using a CNN+RNN model \cite{Zhang2019d}. Another study used transfer learning strategy to improve model generalization for the classification of mental workload \cite{Yin2019a}.

	Multiple kinds of features can be fused to improve assessment performance of mental workload. Gao et al. presented a temporal convolutional block to extract sequential information of EEG. The block orderly consists of a 1D convolution, a rectified linear activation, and a batch normalization. Temporal convolutional blocks and dense layers for spatial feature fusion were combined to form a novel network. Their results showed that this architecture can achieve higher accuracy for fatigue classification, when compared to these networks that replace convolutional block by 1D convolution \cite{Gao2019}. Zhang et al. proposed a two-stream CNN network to learn spectral and temporal features \cite{Zhang2019c}. One stream of CNN was fed by power spectral density topographic maps and the other was fed by topographic maps of amplitude distributions. At the same year (2019), they designed another network for the same propose of learning spectral and temporal features for mental workload classification. In this network, CNN with 3D kernels were first applied to EEG cubes, then extracted features from CNN were flatten to 1D vectors and fed to a bidirectional LSTM for further processing and classification \cite{Zhang2019d}. Both models (i.e. two-stream CNN and CNN+LSTM) showed a significant improvement in mental workload classification.

	\subsection{Sleep Stage Classification}
	Sleep stage classification helps us understand the course of sleep to assess sleep quality and diagnose sleep-related disorders. Table \ref{tab_sleep} briefly summarized the characteristics of each sleep stage. With the aid of EEG recording, sleep quality can be assessed objectively. In the processing of sleep quality, sleep staging is a precedent step. To date, deep learning has been applied to sleep staging. For instance, LSTM model was used for sleep stage classification based on a single channel EEG \cite{Dong2018}. CNN+LSTM model was proposed to classify sleep stages \cite{Bresch2018} \cite{Supratak2017} and detect sleep spindles \cite{Kulkarni2019}.

  Sleep consists of a sequence of stages. Therefore, temporal information should be useful for sleep stage classification. Morlet wavelets \cite{Tsinalis2016} and time-frequency representations \cite{Dong2018} \cite{Hartmann2019} were applied to retain temporal information in the extraction of spectral features. These extracted features were then learned by deep learning models for sleep stage classification, showing promising performance. Using the time-frequency representation of EEG, CNN model achieved good performance \cite{Zhang2020}. In another study, the CNN was combined with LSTM to capture both temporal and spatial information for sleep stage classification \cite{Jeon2019}. The CNN was also combined with attention mechanism for sleep stage classification \cite{eldele2021attention}. In contrast to the supervised learning, unsupervised learning can perform with unlabeled data, which is preferable when the data labelling is expensive or very time-consuming. Zhang et al. presented a CNN model with a greedy layer-wise training strategy, in which complex-valued k-means was utilized to train filters used in the convolution with unlabeled EEG data \cite{Zhang2018}. In \cite{Zhang2016}, unsupervised sparse DBN was used to extract features. Subsequent classifiers (e.g., kNN or SVM) performed well on sleep stage classification by using these unsupervised-extracted features. Jaoude et al. demonstrated that a large training data can help validate classification performance. They trained a deep learning model (CNN+RNN) on sleep data from more than six thousand participants and tested on several publicly available datasets. The model achieved as good as humam experts in sleep staging accuracy \cite{jaoude_expert-level_nodate}. Usually, the numbers of samples for each sleep stage are unbalanced. To date, several methods have been proposed to release this issue, including the class-balanced random sampling \cite{Tsinalis2016}, data augmentation \cite{Mousavi2019}, class-balance training set design \cite{Supratak2017}, and synthetic minority oversampling technique \cite{Chriskos2020}.

	\subsection{Others}
	Those studies that cannot be grouped into the above topics are presented in this subsection. A summary table with key information of those studies is prepared (see Table \ref{tab_other_applications}). On the one hand, EEG with deep learning can be used for person identification \cite{Ozdenizci2019}, \cite{Wang2019}, age and gender prediction \cite{Kaushik2019}. On the other hand, it can also be used to decode brain activity related to vision, audio \cite{Huang2018}, and pain \cite{Yu2020}. In a study of image classification \cite{Jiang2019}, LSTM was used to extract EEG features while CNN was used to extract image features. This study claimed that features extracted from EEG could help image classification so that classification performance was improved. In \cite{Zheng2020}, a CNN+LSTM hybrid network was used to extracted visual representations from EEG, and a generative adversarial network was applied to reconstruct images from the learnt EEG representations. Deep learning and EEG were also applied to understand brain functions and structure. These studies aimed to understand functional brain connectivity \cite{Hua2019}, speech laterality \cite{Toraman2019}, as well as memory under specific conditions. For example, Baltatzis et al. investigated the brain’s activity of different people (ever experienced school bullying or not) to different stimuli (2D videos or Virtual Reality) \cite{Baltatzis2017}. Doborjeh et al. used EEG and spiking neural network to decode how the brain react to various commercial brands (locally familiar or not) \cite{Doborjeh2018}. Arora et al. studied the memory loss after seizure surgery \cite{Arora2018}.

	\section{Discussion}
	In this survey, we reviewed the researches of deep learning in EEG for the last ten years, which is a critical period for the development of deep learning used in EEG. An introduction about deep learning in EEG was first presented in the first section. Subsequently, we presented classical methods of artifacts removal which is an important step in EEG processing. We detailed prevalent deep learning models, followed by the comprehensive reviews on different applications that used deep learning to process and classify EEG signals. These applications were categorised into several topics for presentation. The increase in the number of published papers suggested that the research of deep learning in EEG are expanding over time. Although remarkable achievements were obtained, challenges and limitations still exist, which need to be addressed. We discuss them below and provide our perspectives.

  The performance of deep learning-based classification should be further improved. Although the published papers showed the advantages of deep learning in EEG classification and demonstrated that deep learning is superior to conventional methods, the performance is much lower compared to the performance achieved by deep learning in image or speech classification \cite{rawat2017deep}, \cite{nassif2019speech}. The reasons for the lower performance are mainly due to two aspects: EEG signal itself and deep learning models. On the one hand, EEG signal is non-stationary and much variable over time, which makes the extraction of robust features difficult. An effective solution for this problem is to partition continuous EEG signal into short segments, which can be seen as a stationary signal. However, this is only an approximation but not a final solution. When performing cross-subject classification or cross-session classification, EEG over subjects or sessions is largely variable, making the above problem more dominant. On the other hand, most deep models are originally proposed to process other signals (e.g., images) rather than EEG. Although certain adaptions of the models have been done, the performance is still not ideal because of mismatch between the models and EEG characteristics. Taking CNN as an example, it is more suitable for image processing. Raw images can be directly fed into the CNN. However, this is not the case when applying to EEG signals. Although we have seen some studies, in which raw EEG was fed into CNN directly without pre-processing, it is not mainstream. The mainstream is still to pre-process EEG before feeding into a deep learning model because the pre-processing is very effective for removing noises to improve signal-to-noise ratio. Another advantage of the pre-processing step is that EEG data can be transformed into other representations and/or reorganised to facilitate the following processing in the deep learning model. For instance, spectral power density is one of the most widely used feature for EEG signal. Without a separate pre-processing step, this kind of feature cannot be obtained because temporal EEG signal cannot be transformed into spectral domain within the deep learning model.

Available data size in EEG studies is significantly smaller than that available in image or speech studies \cite{rawat2017deep}, \cite{nassif2019speech}. As we know, the deep learning model requires extensive training and a large data size can benefit model training to a great extent. Compared to the millions of training data in image or speech recognition, the scale of training data is much less in EEG classification, only from tens, hundreds, or at most thousands of participants. One potential solution for the lack of EEG data in the model training is the use of transfer learning. Deep learning model can be trained by the data which are not collected at the moment and the trained model can be used for recognition or classification on the new collected EEG data after fine-tuning or even without fine-tuning \cite{Lu2017},  \cite{Hang2019}, \cite{Zhao2019}. Unlike image classification, for which there are mature existing pre-trained models (e.g., ImageNet pre-trained VGG model), there is no publicly available pre-trained model for EEG classification. If VGG model is directly applied to EEG, reorganization of EEG has to be done in order to meet the input data format of VGG model. This reorganization might lead to information loss and give detrimental effect on the EEG classification. In addition, there is no idea how well a model trained on images can be tuned to classify EEG signal.

 Based on the effectiveness comparison of transfer learning, greater performance improvement was observed in image classification compared to EEG classification. This might be due to the lack of effective training framework and strategies that are suitable for transferring EEG patterns. There was an attempt to transfer the model trained on images to EEG classification \cite{Xu2019a}. This transferring is across distinct modalities. It is likely to have a better performance when transferring across relevant modalities. As we know, there are different modalities (e.g., functional near-infrared spectroscopy (fNIRS) and EEG) that can be used to measure underlying brain activity. A deep learning model can be trained on one modality and then fine-tuned by the other modality to classify signals of that modality. Or, different modalities can be used together to train a deep learning model so that the training can be benefited from the complementary information existing in the different modalities. It is a fusion of modalities. It has been seen that classification performance was elevated by feature fusion in the case of using conventional classifiers \cite{Ho2019}. The fusion could be done at the different stages of the classification process (e.g., at the beginning of initial feature fusion or at the later stage of decision fusion \cite{harvy2018performance}, \cite{pei2020eeg}). Wu et al. utilized both EEG and Electrooculogram (EOG) to classify the level of vigilance by fusing the features extracted from EEG and EOG \cite{wu2020multimodal}. In the future, more extensive research should be carried out to elevate the development of fusion in deep learning models. Especially, to address how to effectively fuse multiple modalities in deep learning models for neurophysiological signal classification and analysis. Of course, collecting adequate data is a straightforward solution for the lack of EEG data. However, this results in new issues, such as cost increase and time delay. If data collection involves different institutes, extra communication effort should be paid to coordinate the data collection. Meanwhile, computation demand will be increased with the increase of data size, which requires to upgrade computational hardware or replace with the new generation hardware (e.g., central processing unit (CPU) and graphics processing unit (GPU)). As mentioned in \cite{li2020cognitive}, cloud computing service is an effective way to share hardware resources so that the hardware cost in individual institutes will be reduced. Using the cloud computing service, data protection and privacy have to be considered, especially for clinical data.

 When applying a deep learning model to EEG, we need to adapt the deep learning model in compliance with the characteristics of EEG. For example, how to arrange the input data or how to set kernel size should be considered. EEG signal is usually not directly used and commonly transformed before feeding into a deep learning model. There are strong relationships among temporal domain, spectral domain, and spacial domain. It is important these relationships should be kept as much as possible when arranging the input data. When EEG channels are stacked along a dimension, their spacial layout is distorted. In this case, kernels, such as square kernel, that usually-used in image recognition are no longer effective for EEG classification. A column kernel (covering all channels) is a better choice, which has been supported by the study in \cite{Goh2018}. Further, Wang et al. extended the column kernel by considering brain anatomic structure to develop multiple kernels with the sizes matching brain region sizes, achieving a better performance in schizophrenia identification compared to the usually-used kernels, such as square kernel \cite{wang2020multikernel}.

We believe deep learning models should be changed to be more flexible. The trained model can be adapted dynamically in real-time as needed. This is not limited to dynamic parameter tuning. Ideally, model architecture can also be adjusted when needed. Also, we hope the newly-developed deep learning model could perform multiple tasks at the same time in the future. Please see the detailed description in \cite{li2020thoughts}.

Apart from the purposes of deep learning-based EEG classification, deep learning may also be a useful tool to reveal neural mechanisms of the brain. When a deep learning model achieves a satisfactory classification performance, it captures essential differences existing between the classes. Therefore, we can look at what information the deep learning model focuses on to roughly infer the underlying associated brain activity. For example, Goh et al. presented spatial distribution of brain activations associated with lower limb movements by probing into the model of spatio-spectral representation learning \cite{Goh2018}. We expect that advanced deep learning models developed in the future could reversely decompose EEG signal back into the representation in the brain to reveal underlying brain mechanisms. It is unrealistic at the current stage, but paying efforts to make progress towards to this target.

  A prominent advance we need to mention is the EEGNet \cite{Lawhern2018}, which is proven effective for different BCI paradigms. Another promising model is SincNet, which was initially proposed for speaker recognition and also well for the classification of EEG signal \cite{Zeng2019}. New deep learning architectures, such as capsule network \cite{wang2020multikernel}, are also required to enhance the chance of success of EEG applications.

  Lastly, a mix of different deep learning units has been increasingly seen, which integrates the characteristics of these units to benefit data learning. Because there is not definite guidance to set optimal deep learning architecture (e.g., model depth and model width) currently, model complexity might be considered to determine the model architecture. The model should have enough capacity for learning information in accordance with classification tasks while its complexity should be kept as low as possible to minimize computational cost.

	\section{Conclusion}
	Our survey is a glimpse of what have been done for the deep learning in EEG over the past ten years. There are still many researches currently on-going at laboratories and hospitals, dealing with challenges we mentioned above and beyond. We hope that our survey can provide the researchers who are working in this field with a summary and facilitate their researches.

	\bibliographystyle{IEEEtran}
	\bibliography{second.bib}

	\clearpage
\onecolumn

\begin{table*} 
	\caption{The Abbreviations in This Survey}
	\label{tab_abbr}
	\centering
	\begin{tabular}{ll}  
		\toprule    
		Abbreviation & Full Name \\
		\midrule
		AD  & Alzheimer's Disease   \\
		ADHD  & Attention Deficit Hyperactivity Disorder    \\
		AE  & Autoencoder   \\
		BCI & Brain-Computer Interface    \\
		CAM-ICU & Confusion Assessment Method for the ICU   \\
		CapsNet & Capsule Network   \\
		CJD & Creutzfeldt-Jakob Disease   \\
		CNN & Convolutional Neural Network    \\
		DBCS  & Deep Blind Compresed Sensing    \\
		DBN & Deep Belief Network   \\
		DMCCA& Deep Multiset Canonical Correlation Analysis\\
		DN-AE-NTM&Deep Network Autoencoder Neural Turing Machine\\
		DPN & Deep Polynomial Network   \\
		DTI & Diffusion Tensor Imaging    \\
		DWT & Discrete Wavelet Transformation   \\
		EEG & Electroencephalogram    \\
		ELM & Extreme Learning Machine    \\
		EOG & Electrooculogram \\
		ESN & Echo State Network    \\
		fMRI  & functional Magnetic Resonance Imaging   \\
		fNIRS & functional Near-Infrared Spectroscopy   \\
		GPED  & Generalized Periodic Epileptiform Discharge   \\
		GRU & Gated Recurrent Unit    \\
		HC  & Healthy Controls    \\
		IED & Iterictal Epileptiform Discharge    \\
		kNN & k-Nearest Neighbor    \\
		LSTM  & Long Short-Term Memory    \\
		MCI & Mild Cognitive Impairment   \\
		MI  & Motor Imagery   \\
		MLP & Multilayer Perceptron   \\
		NREM  & Non-Rapid Eye Movement    \\
		OFS & Operator Functional States    \\
		PCA & Principal Component Analysis    \\
		PLED  & Periodic Lateralized Epileptiform Discharge   \\
		RASS  & Richmond Agitation-Sedation Scale   \\
		RBM & Restricted Boltzmann Machine    \\
		REM & Rapid Eye Movement    \\
		RNN & Recurrent Neural Network    \\
		RPD & Rapidly Progressive Dementia    \\
		RSVP  & Rapid Serial Visual Presentation    \\
		SAE & Stacked Autoencoder   \\
		SAN & Subject Adaption Network    \\
		SNN & Spiking Neural Network    \\
		SNR & Signal-to-Noise Ratio   \\
		SRP & Sparse Random Projection    \\
		SSRL  & Spatio-Spectral Represemtation Learning   \\
		SSVEP & Steady State Visually Evoked Potentials     \\
		SVM & Support Vector Machine    \\
		
		\bottomrule  
	\end{tabular}
\end{table*}

\begin{landscape}
	\tiny	
  
	\clearpage
	
	 \begin{spacing}{2.8}

	\begin{longtable}{p{5cm}p{5cm}p{7cm}}
		\caption{Typical Methods for Artifacts Removal} \\
		\label{tab_artifacts}\\
		\toprule			
		Methods	&	Target Artifacts	&	Property	\\
		\midrule
		Notch Filter	&	Line Noise	&	Signal distortion in specific frequencies	\\
		Band-Pass Filter	&	Artifacts concentrated on a particular frequency band	&	Preclude certain frequency signals	\\
		Independent Component Analysis	&	Ocular and muscular noise removal	&	Decompose channels into independent components	\\
		Reject Contaminated Data Segments	&	Ocular noise, muscular noise etc., which are difficultly mitigated	&	Reject gross eye movement and occasional recording artifacts	\\
		Wavelet Transformation Analysis	&	Ocular and muscular noise removal	&	Signals are reconstructed based on the corrected coefficient	\\
		Common Average Reference	&	Artifacts equivalently affect all channels	&	Amplitudes can be overall reduced	\\
		Z-Score Calculation	&	Noisy channels or time periods	&	Generates zero-mean data with unitary variance	\\
		Denoise AutoEncoder	&	General Noises	&	Denoise in an unsupervised manner	\\
		\bottomrule
	\end{longtable}
	\clearpage
	\begin{longtable}[r]{p{3cm}p{2.3cm}p{1.7cm}p{7.2cm}p{8cm}}
		\caption{Key Information of Papers about Brain-Computer Interface} \\
		\label{tab_bci}\\
		\toprule
		Authors			&	Models	&	Paradigms	&	Classes	&	Data (Private/Public: No. of Participants, No. of Channels, Sampling Rate)	\\
		\midrule
		Ma et al. 2020  \cite{Ma2020} & CNN & MI  & Rest, Right Hand, and Right Elbow   & Private: 25 Participants, 64 Channels, 1000 Hz  \\
		Zhang et al.  2019  \cite{Zhang2019a} & CNN & MI  & Left, Right Hand  & BCI Competition II Dataset III  \\
		Xu et al. 2019  \cite{Xu2019a}  & CNN & MI  & Left, Right Hand  & BCI Competition IV Dataset 2b \\
		Zhu et al.  2019  \cite{Zhu2019}  & CNN & MI  & Left, Right Hand  & \tabincell{l}{1. Private: 25 Participants, 15 Channels, 1000 Hz \\
			2. BCI Competition IV Dataset 2b} \\
		Lu et al. 2017  \cite{Lu2017} & DBM & MI  & Left, Right Hand  & BCI Competition IV Dataset 2b \\
		Chiarelli et al.  2018  \cite{Chiarelli2018}  & DNN & MI  & Left, Right Hand  & Private: 15 Participants, 128 Channels, 250 Hz  \\
		Tayeb et al.  2019  \cite{Tayeb2019}  & CNN, LSTM, CNN+LSTM & MI  & Left, Right Hand  & \tabincell{l}{1. Private: 20 Participants, 32 Channels, 256 Hz\\
			2. BCI Competition IV Dataset 2b} \\
		Dai et al.  2019  \cite{Dai2019}  & CNN+AE  & MI  & Left, Right Hand  & BCI Competition IV Dataset 2b \\
		Ha et al. 2019  \cite{Ha2019} & CapsNet & MI  & Left, Right Hand  & BCI Competition IV Dataset 2b \\
		Shi et al.  2019  \cite{Shi2019}  & CNN & MI  & Left, Right Hand  & Private: - Participants, 118 Channels, - Hz \\
		Wang et al. 2018  \cite{Wang2018} & CNN, LSTM & MI  & Left, Right Hand  & Private: 14 Participants, 11 Channels, 256 Hz \\
		Tabar et al.  2017  \cite{Tabar2017}  & CNN, SAE, CNN+SAE & MI  & Left, Right Hand  & \tabincell{l}{1. BCI Competition II Dataset III\\
			2. BCI Competition IV Dataset 2b} \\
		Amin et al. 2019  \cite{Amin2019a}  & CNN & MI  & Left Hand, Right Hand, Feet, and Tongue & \tabincell{l}{1. High Gamma Dataset \cite{schirrmeister2017deep}\\
			2. BCI Competition IV Dataset 2a} \\
		Amin et al. 2019  \cite{Amin2019} & CNN, MLP, AE  & MI  & Left Hand, Right Hand, Feet, and Tongue & \tabincell{l}{1. BCI Competition IV Dataset 2a\\
			2. High Gamma Dataset \cite{schirrmeister2017deep}} \\
		Li et al. 2019  \cite{Li2019a}  & CNN & MI  & Left Hand, Right Hand, Feet, and Tongue & BCI Competition IV Dataset 2a \\
		Hassanpour et al. 2019  \cite{Hassanpour} & DBN, SAE  & MI  & Left Hand, Right Hand, Feet, and Tongue & BCI Competition IV Dataset 2a \\
		Zhang et al.  2019  \cite{Zhang2019f} & CNN+LSTM  & MI  & Left Hand, Right Hand, Feet, and Tongue & BCI Competition IV Dataset 2a \\
		She et al.  2018  \cite{She2019}  & ELM & MI  & Left Hand, Right Hand, Feet, and Tongue & BCI Competition IV Dataset 2a \\
		Uribe et al.  2019  \cite{Uribe2019}  & ELM & MI  & Left Hand, Right Hand, Feet, and Tongue & BCI Competition IV Dataset 2a \\
		Lei et al.  2019  \cite{Lei2019}  & MMDPN & MI  & Idle, Preparation, Walking Imagery, and Restoration & Private: 9 Participants, 32 Channels, 512 Hz  \\
		Duan et al. 2017  \cite{Duan2017} & ELM & MI  & Cortical Positivity and Negativity  & BCI Competition II Dataset Ia \\
		Alazrai et al.  2019  \cite{Alazrai2019}  & CNN & MI  & Rest, Grasp-Related (Small Diameter, Lateral, and Extension-Type), Wrist-Related (Ulnar/Radial Deviation. Flexion/Extension), Fingers-Related ( Flexion and Extension of The Index, The Middle, The Ring, The Little, and The Thumb Finger) & Private: 22 Participants (18 Able-Bodied and 4 with Transradial Amputations), 16 Channels, 2048 Hz  \\
		Hang et al. 2019  \cite{Hang2019} & CNN & MI  & \tabincell{l}{1. Right Hand, Foot\\
			2. Left Hand, Right Hand, Feet, and Tongue} & \tabincell{l}{1. BCI Competition III Dataset IVa\\
			2. BCI Competition IV Dataset IIa}  \\
		Yang et al. 2018  \cite{Yang2018a}  & CNN+LSTM  & MI  & \tabincell{l}{1. Left Hand, Right Foot\\
			2. Left, Right Hand\\
			3. Left Hand, Tongue} & \tabincell{l}{1. Private: 6 Participants, 64 Channels, 500 Hz\\
			2. BCI Competition III Dataset -\\
			3. BCI Competition IV Dataset -}  \\
		Zhao et al. 2019  \cite{Zhao2019} & CNN & MI  & \tabincell{l}{1. Left Hand, Right Hand, Feet, and Tongue\\
			2. Left, Right Hand\\
			3. Elbow Flexion/Extension, Forearm Supination/Pronation, Hand Open/Close}  & \tabincell{l}{1. BCI Compeition IV Dataset 2a\\
			2. BCI Compeition IV Dataset 2b\\
			3. From Ofner et al., 15 Participants, 61 Channels, 512 Hz} \\
		Wu et al. 2019  \cite{Wu2019} & CNN & MI  & \tabincell{l}{1. Left Hand, Right Hand, Feet, and Tongue\\
			2. Left, Right Hand}  & \tabincell{l}{1. BCI Competition IV Dataset 2a\\
			2. BCI Competition IV Dataset 2b\\
			3. High Gamma Dataset \cite{schirrmeister2017deep}} \\
		Majidov et al.  2019  \cite{Majidov2019}  & CNN & MI  & \tabincell{l}{1. Left Hand, Right Hand, Feet, and Tongue\\
			2. Left, Right Hand}  & \tabincell{l}{1. BCI Competition IV Dataset 2a\\
			2. BCI Competition IV Dataset 2b} \\
		Li et al. 2019  \cite{Li2019c}  & CNN & MI  & \tabincell{l}{1. Left Hand, Right Hand, Feet, and Tongue\\
			2. Left Hand, Right Hand, Feet, and Rest} & \tabincell{l}{1. BCI Competition IV Dataset 2a\\
			2. High Gamma Dataset \cite{schirrmeister2017deep}} \\
		Dose et al. 2018  \cite{Dose2018} & CNN & MI  & Left/Right Fist or Both Fists/Both Feet & EEG Motor Movement/MI Dataset \\
		Tang et al. 2019  \cite{Tang2019} & DBN & MI  & Left, Right Hand  & Private: 7 Participants, 14 Channels, 128 Hz  \\
		Xu et al. 2018  \cite{Xu2019} & CNN & MI  & \tabincell{l}{1. Left, Right Hand\\
			2. Left Hand, Right Hand, Feet, and Tongue} & \tabincell{l}{1. BCI Competition II Dataset III\\
			2. BCI Competition IV Dataset 2a} \\
		Kwon et al. 2020  \cite{kwon_subject-independent_2020}  & CNN & MI  & Left and Right Hnad & Private: 54 Participants, 62 Channels, 1000 Hz  \\
		Mammone et al.  2020  \cite{mammone_deep_2020}  & CNN & MI  & Elbow Flexion/Extension, Forearm Supination/Pronation, Hand Open/Close, Resting & BNCI Horizon Dataset  \\
		Zhang et al.  2020  \cite{zhang_motor_2020} & CNN+LSTM  & MI  & \tabincell{l}{1. Left/Right Fist Open and Close\\
			2. Left hand, right hand, feet, and tongue} & \tabincell{l}{1. PhysioNet Dataset\\
			2. BCI Competition IV Dataset 2a} \\
		Chen et al. 2020  \cite{chen_deep_2020} & CNN & MI  & \tabincell{l}{1. Left hand, right hand, feet, and tongue\\
			2. Right hand and feet} & \tabincell{l}{1. BCI Competition IV Dataset 2a\\
			2. SMR-BCI Dataset} \\
		Jeong et al.  2020  \cite{jeong_brain-controlled_2020} & CNN+LSTM  & Reaching Movements and MI & Left, Right, Forward, Backward, Up, and Down  & Private: 15 Participants, 64 Channels, 1000 Hz  \\
		Ding et al. 2015  \cite{Ding2015} & ELM & - & Cortical Positivity and Negativity  & BCI Competition II Dataset Ia \\
		Ma et al. 2017  \cite{Ma2017} & DBN & mVEP  & Target Stimulus Signal and The Standard Stimulus Signal & Private: 11 Participants, 10 Channels, 1000 Hz  \\
		Gao et al.  2015  \cite{Gao2015}  & ANN & P300  & P300 and Non-P300 & Private: 5 Participants, 32 Channels, 2048 Hz \\
		Kundu et al 2019  \cite{Kundu2019}  & SAE & P300  & P300 and Non-P300 & \tabincell{l}{1. BCI Competition II Dataset IIb\\
			2. BCI Competition III Dataset II\\
			3. BNCI Horizon Dataset}  \\
		Kshiragar et al.  2019  \cite{Kshirsagar2019} & SAE, CNN & P300  & P300 and Non-P300 & Private: 10 Participants, 16 Channels, 500 Hz \\
		Liu et al.  2018  \cite{Liu2018}  & CNN & P300  & P300 and Non-P300 & \tabincell{l}{1. BCI Competition III Dataset II\\
			2. BCI Competition II Dataset IIb}  \\
		Farahat et al.  2019  \cite{Farahat2019}  & CNN & P300  & P300 and Non-P300 & Private: 19 Participants, 29 Channels, 508.63 Hz  \\
		Solon et al.  2019  \cite{Solon2019}  & CNN & P300  & P300 and Non-P300 & Private: 67 Participants, 64 Channels, - Hz \\
		Vareka et al. 2017  \cite{Vareka2017} & SAE & P300  & P300 and Non-P300 & Private: 25 Participants, 19 Channels, 1000 Hz  \\
		Morabbi et al.  2018  \cite{Morabbi2019}  & DBN & P300  & P300 and Non-P300 & EPFL BCI Dataset  \\
		Ditthapron et al. 2019  \cite{Ditthapron2019} & CNN+LSTM+AE & P300  & P300 and Non-P300 & \tabincell{l}{1. From Citi et al. \cite{citi2010documenting}, 12 Participants, 64 Channels, 2048 Hz\\
			2. BCI Competition III Dataset II\\
			3. From Schreuder et al. \cite{schreuder2010new}, 10 Participants, 60 Channels, 240 Hz\\
			4. From Acqualagna et al. \cite{acqualagna2013gaze}, 13 Participants, 63 Channels, 250 Hz\\
			5. EEG Database Data Set/UCI EEG Dataset\\
			6. From Treder et al. \cite{treder2014decoding}, 11 Participants, 63 Channels, 200 Hz}  \\
		Lawhern et al.  2018  \cite{Lawhern2018}  & CNN & P300, MI, etc. & \tabincell{l}{1. P300 and Non-P300\\
			2. Correct and Incorrect\\
			3. The Left Index, Left Middle, Right Index, and Right Middle Finger\\
			4. Left Hand, Right Hand, Feet, and Tongue} & \tabincell{l}{1. Private: 15 Participants, 64 Channels, 512 Hz\\
		2. BCI Challenge\\
		3. Private: 13 Participants, 256 Channels, 1024 Hz\\
		4. BCI Competition IV Dataset 2a  }\\
		Boloukian et al.  2020  \cite{Boloukian2020}  & DN-AE-NTM & P300, MI, etc. & \tabincell{l}{1. P300 and Non-P300\\
			2. Alcoholic and Control\\
			3. Left/Right Fist or Both Fists/Both Feet} & \tabincell{l}{1. From Hoffmann et al. \cite{hoffmann2008efficient}, 9 Participants (5 with disablement and 4 able-bodied), - Channel, - Hz\\
			2. EEG Database Data Set/UCI EEG Dataset\\
			3. EEG Motor Movement/Imagery Dataset}  \\
		Pei et al.  2018  \cite{Pei2018}  & SAE & Reaching Movements  & Left, Central and Right & Private: 5 Participants, 32 Channels, 256 Hz  \\
		Chen et al. 2019  \cite{Chen2019c}  & CNN & RSVP  & Target and Non-Target & From Touryan et al. \cite{touryan2013translation}, 10 Participants, 64 Channels, 512Hz  \\
		Manor et al.  2015  \cite{Manor2015}  & CNN & RSVP  & Target and Non-Target & Private: 15 Participants, 64 Channels, 256 Hz \\
		Manor et al.  2016  \cite{Manor2016}  & CNN & RSVP  & Target and Non-Target & Private: 15 Participants, 64 Channels, 256 Hz \\
		Nguyen et al. 2019  \cite{Nguyen2019} & CNN & SSVEP & 6.67, 7.5, 8.57, 10, and 12 Hz  & Private: 8 Participants, 1 Channel, 128 Hz  \\
		Liu et al.  2020  \cite{Liu2020}  & DMCCA & SSVEP & 6, 8, 9, and 10 Hz  & Private: 10 Participants, 8 Channels, 250 Hz  \\
		Waytowich et al.  2018  \cite{Waytowich2018}  & CNN & SSVEP & 12 SSVEP Stimuli Flashed at Frequencies Ranging
		from 9.25 Hz To 14.75 Hz in Steps of 0.5 Hz & From Nakanishi et al. \cite{nakanishi2015comparison}, - Participants, - Channel, 2048 Hz  \\				
		\bottomrule
		\multicolumn{5}{l}{'-' indicates that the information is unavailable} \\
	\end{longtable}
	\clearpage

	\begin{longtable}{p{3cm}p{1.7cm}p{3cm}p{6cm}p{9cm}}
		\caption{Key Information of Papers about Disease Detection} \\
		\label{tab_disease}\\
		\toprule	
		Author			&	Models	&	Categories	&	Classes	&	Data (Private/Public: No. of Participants, No. of Channels, Sampling Rate)	\\
		\midrule
		Doborjeh et al. 2016  \cite{Doborjeh2016} & SNN & Addiction & Healthy, Addiction Treated, and Addiction Not Treated Subjects & Private: 74 Participants, 26 Channels, - Hz \\
		Ieracitano et al. 2019  \cite{Ieracitano2019} & CNN & Alzheimer's Disease & \tabincell{l}{1. AD vs. HC, AD vs. MCI, MCI vs. HC\\
			2. AD, MCI, and HC} & Private: 189 Participants (63 AD, 63 MCI, 63 HC), 19 Channels, 1024 Hz  \\
		Bi et al. 2019  \cite{Bi2019} & DBN & Alzheimer's Disease & \tabincell{l}{1. AD, HC, and MCI\\
			2. Identification: determine EEG spectral image come from which person\\
			3. Verification: wheather two EEG spectral images come from the same person}  & Private: 12 Participants (4 HC, 4 MCI, and 4 AD), 64 Channels, 500 Hz \\
		Morabito et al. 2016  \cite{Morabito2017} & SAE, MLP  & Alzheimer's Disease & CJD/RPD, CJD/HC, and CJD/AD & Private: 76 Participants, 19 Channels, - Hz \\
		Hayase et al. 2019  \cite{Hayase} & MLP & Anaesthesia & - & Private: 30 Participants, - Channels, 128 hZ  \\
		Liu et al.  2019  \cite{Liu2019}  & CNN & Anaesthesia & Anesthetic Ok, Deep, and Light  & Private: 50 Participants, - Channel, - Hz \\
		Park et al. 2020  \cite{park_real-time_2020}  & CNN & Anesthesia  & - & VitalDB \\
		Kim et al.  2018  \cite{Kim2018a} & CNN, LSTM, DNN  & Brain Disease & \tabincell{l}{1. Normal and Dementia\\
			2. Normal and Alcoholism} & EEG Database Data Set/UCI EEG Dataset \\
		Chen et al. 2019  \cite{Chen2019e}  & CNN & Children with ADHD  & Adhd and Controls & Private: 107 Participants (50 Children with ADHD and 57 Controls), 128 Channels, 1000 Hz  \\
		Chen et al. 2019  \cite{Chen2019d}  & CNN & Children with ADHD  & Adhd and Controls & Private: 107 Participants (50 Children with ADHD and 57 Controls), 62 Channels, 1000 Hz \\
		Boshra et al. 2019  \cite{Boshra2019} & CNN & Concussion  & Normal and Concussion & Private:  54 Participants (26 with Concussion and 28 Controls), 64 Channels, 512 Hz  \\
		Sun et al.  2019  \cite{Sun2019a} & CNN+LSTM  & Consciousness and Delirium Tracking & \tabincell{l}{1. Rass: -5, -4, -3, -2, -1, 0\\
			2. Cam-Icu: 0, 1} & Private: 295 Participants (174 for RASS and 121 for CAM-ICU), 4 Channels, 250 Hz  \\
		Ay et al. 2019  \cite{Ay2019} & CNN+LSTM  & Depression  & Normal and Depression & From Acharya et al. \cite{acharya2018automated}, 30 Participants (15 Depressed and 15 Normal), 1 Channel (FP1-T3, FP2-T4), 256 Hz \\
		Acharya et al.  2018  \cite{Acharya2018}  & CNN & Depression  & Depression and Normal & Private: 30 Participants (15 Deoressed and 15 Normal), FP1-T3 and FP2-T4 Channel, 256 Hz  \\
		Li et al. 2019  \cite{Li2019} & CNN & Depression  & Depression and Normal & Private: 28 Participants (14 Deoressed and 14 Normal), 16 Channels, 250 Hz  \\
		Mumtaz et al. 2019  \cite{Mumtaz2019} & CNN, CNN+LSTM & Depression  & Depression and Normal & Private: 63 Participants (33 Deoressed and 30 Normal) \\
		Zhu et al.  2019  \cite{Zhu2019a} & MDAE  & Depression  & Mild Depression and Normal  & Private: 51 Participants (24 Mild Deoression and 27 Normal), 16 Channels, 250 Hz  \\
		Bouallegue et al. 2020  \cite{bouallegue_dynamic_2020}  & RNN+CNN & Autism and Epilepsy & \tabincell{l}{1. Normal and Autistic\\
			2. Normal and Seizure}  & \tabincell{l}{1. Private: 19 Participants (10 normal and 9 autistic), 16 Channels, 256 Hz\\
			2. CHB-MIT Scalp EEG database\\
			3. From Andrzejak et al.\cite{andrzejak2001indications}, 10 participants (5h healthy and 5 epileptic patients)} \\
		Cao et al.  2020  \cite{cao_epileptic_2020} & CNN+ELM & Epilepsy  & \tabincell{l}{1. Seizure/Non-Seizure\\
			2. Interictal, Preictal, Ictal\\
			3. Interictal, Three Preictal States, Ictal}  & \tabincell{l}{1. CHB-MIT Scalp EEG database\\
			2. Private: 10 Participants, 18 Channels, 256 Hz} \\
		Daoud et al.  2020  \cite{Daoud_deep_2020}  & CNN+AE+MLP  & Epilepsy  & Focal and Non-Focal & \tabincell{l}{1. From Andrzenak et al.\cite{andrzejak2012nonrandomness}, 5 epileptic patients\\
			2. From Andrzejak et al.\cite{andrzejak2001indications}, 10 participants (5h healthy and 5 epileptic patients)} \\
		Tsiouris et al. 2018  \cite{Tsiouris2018} & LSTM  & Epilepsy  & Preictal and Interictal & CHB-MIT Scalp EEG Database  \\
		Yuan et al. 2019  \cite{Yuan2019a}  & AE  & Epilepsy  & Ictal and Non-Ictal & CHB-MIT Scalp EEG Database  \\
		Karim et al.  2019  \cite{Karim2019}  & SAE & Epilepsy  & Healthy and Epileptic Activiy & From Andrzejak et al. \cite{andrzejak2001indications}, 10 Participants (5 Healthy and 5 Epileptic Patients) \\
		Ullah et al.  2018  \cite{Ullah2018}  & CNN & Epilepsy  & \tabincell{l}{1. Seizure, and Non-Seizure\\
			2. Normal, Interical, and Ictal}  & From Andrzejak et al. \cite{andrzejak2001indications}, 10 Participants (5 Healthy and 5 Epileptic Patients) \\
		San-Segundo et al.  2019  \cite{San-Segundo2019}  & CNN & Epilepsy  & \tabincell{l}{1. Focal and Non-Focal\\
			2. Healthy/Ictal, Ictal/Non-Ictal, Healthy/\\Non-Focal/Ictal, and Healthy/Focal/Ictal}
		& \tabincell{l}{1. The Bern-Barcelona EEG Database\\
			2. Epileptic Seizure Recognition Data Set}  \\
		Wen et al.  2018  \cite{Wen2018}  & CNN+AE  & Epilepsy  & \tabincell{l}{1. Health With Eyes Open/Closed (A, B),\\ Interictal (C, D), and Ictal (E)\\
			2. Epileptic Seizure and Non-Epileptic Seizure} & \tabincell{l}{1. From Andrzejak et al.\cite{andrzejak2001indications}, 10 Participants (5 Healthy and 5 Epileptic Patients)\\
			2. CHB-MIT Scalp Database}  \\
		Acharya et al.  2018  \cite{Acharya2018a} & CNN & Epilepsy  & Noraml, Preictal, and Seizure & From Andrzejak et al. \cite{andrzejak2001indications}, 10 Participants (5 Healthy and 5 Epileptic Patients) \\
		Qiu et al.  2018  \cite{Qiu2018}  & SAE & Epilepsy  & Normal, Interictal, and Ictal & From Andrzejak et al. \cite{andrzejak2001indications}, 10 Participants (5 Healthy and 5 Epileptic Patients) \\
		Turk et al. 2019  \cite{Turk2019} & CNN & Epilepsy  & \tabincell{l}{1. A and B\\
			2. A, B, and E\\
			3. A, C, D, and E\\
			4. A, B, C, D, and E} & From Andrzejak et al. \cite{andrzejak2001indications}, 10 Participants (5 Healthy and 5 Epileptic Patients) \\
		Thara et al.  2019  \cite{Thara2019}  & LSTM  & Epilepsy  & \tabincell{l}{1. Seizure and Non-Seizure\\
			2. Preictal, Interictal, and Ictal }  & From Bonn University, 500 Participants (missing detial) \\
		Sayeed et al. 2019  \cite{AbuSayeed2019}  & DNN & Epilepsy  & \tabincell{l}{1. Normal and Ictal\\
			2. Normal. Interictal, and Ictal} & From Andrzejak et al.\cite{andrzejak2001indications}, 10 Participants (5 Healthy and 5 Epileptic Patients)  \\
		Hosseini et al. 2017  \cite{Hosseini2017} & CNN, SAE  & Epilepsy  & Interictal, and Preictal
		& \tabincell{l}{1. Private: 9 Participants, 70 Channels, 1000 Hz\\
			2. From Upenn and the Mayo Clinic \cite{stead2010microseizures} \cite{brinkmann2009large}, 2 Participants, 15 Channels, 5000 Hz}  \\
		Hussein et al.  2019  \cite{Hussein2019}  & LSTM  & Epilepsy  & \tabincell{l}{1. Normal and Seizure\\
			2. Normal, Inter-Ictal, and Ictal\\
			3. Health With Eyes Open/Closed (A, B), Interictal (C, D), and Ictal (E)} & From Andrzejak et al. \cite{andrzejak2001indications}, 10 Participants (5 Healthy and 5 Epileptic Patients) \\
		Abdelhameed et al.  2019  \cite{Abdelhameed2019}  & CNN+AE  & Epilepsy  & \tabincell{l}{1. Normal and Ictal\\
			2. Normal. Interictal, and Ictal} & From Andrzejak et al. \cite{andrzejak2001indications}, 10 Participants (5 Healthy and 5 Epileptic Patients) \\
		He et al. 2019  \cite{He2019} & CNN & Epilepsy  & Five Classes: Health With Eyes Open/Closed (A, B), Interictal (C, D), and Ictal (E) & From Andrzejak et al. \cite{andrzejak2001indications}, 10 Participants (5 Healthy and 5 Epileptic Patients) \\
		Cao et al.  2019  \cite{Cao2019}  & CNN+LSTM  & Epilepsy  & Iic Patterns and Others & From MGH, over 2500 Participants, 20 Channels, - Hz \\
		Akut  2019  \cite{Akut2019} & CNN & Epilepsy  & \tabincell{l}{1. Normal and Ictal\\
			2. Normal. Interictal, and Ictal} & From Andrzejak et al. \cite{andrzejak2001indications}, 10 Participants (5 Healthy and 5 Epileptic Patients) \\
		Emami et al.  2019  \cite{Emami2019}  & CNN & Epilepsy  & Seizure and Non-Seizure & \tabincell{l}{1. Private: 8 Participants, 19 Channels, 1000 Hz\\
			2. Private: 16 Participants, 19 Channels, 500 Hz} \\
		Daoud et al.  2019  \cite{Daoud2019}  & MLP, CNN, LSTM, SAE & Epilepsy  & Interictal and Preictal & CHB-MIT Scalp EEG Database  \\
		Tian et al. 2019  \cite{Tian2019} & CNN & Epilepsy  & Seizure and Non-Seizure & CHB-MIT Scalp EEG Database  \\
		Wei et al.  2018  \cite{Wei2018}  & CNN & Epilepsy  & Interictal, Preictal, and Ictal & Private: 13 Participants, 22 Channels, 500 Hz \\
		Antoniades et al. 2017  \cite{Antoniades2017} & CNN & Epilepsy  & IED and Non-IED & Private: 18 Participants, 20 Channels, 200 Hz \\
		Baloglu et al.  2019  \cite{Baloglu2019}  & CNN+LSTM  & Epilepsy  & Normal/Ictal, Interictal/Ictal, Normal/Epilepsy, Nonictal/Ictal, Normal/Interictal/Ictal  & From Andrzejak et al. \cite{andrzejak2001indications}, 10 Participants (5 Healthy and 5 Epileptic Patients) \\
		Oshea et al.  2019  \cite{oshea_neonatal_2020}  & CNN & Epilepsy  & Seizure and Non-Seizure & \tabincell{l}{1. Private: 18 Participants, 8 Channels, 256 Hz\\
			2. Helsinki Dataset}  \\
		Vrbancic et al. 2018  \cite{Vrbancic2018} & CNN & Motor Impairment Neural Disorders & Normal and Motor Impairments  & CSU BCI collection  \\
		Jansen et al. 2018  \cite{Jansen2018} & ANN & Obstructive Sleep Apnea & OSA Patients and Controls & From Klosch et al. \cite{klosh2001siesta}, 247 Participants (50 Patients and 197 Controls), 6 Channels, - Hz  \\
		Jonas et al.  2019  \cite{Jonas2019}  & CNN & Outcome Prediction after Cardiac Arrest & Favorable and Unfavorable Outcome & Private: 267 Participants, 19 Channels, 250 Hz  \\
		Hofmejer et al. 2018  \cite{Tjepkema-Cloostermans2019}  & CNN & Outcome Prediction in Postanoxic Coma & Good and Poor & Private: 456 Participants, - Channels, - Hz \\
		Amin et al. 2019  \cite{Amin2019b}  & CNN & Pathology & Normal and Pathology  & TUH Abnormal EEG Dataset  \\
		Ruffini et al.  2019  \cite{Ruffini2019}  & CNN & REM Behavior Disorder (RBD)  & \tabincell{l}{1. HC and Parkinson’S Disease (PD)\\
			2. HC+ RBD Vs. PD+Dementia with Lewy Bodies(DLB)} & Private: 206 Participants (121 with Idiopathic RBD), 14 Channels, 256 Hz  \\
		Naira et al.  2019  \cite{Naira2019}  & CNN & Schizophrenia & Normal and Schizophrenia  & From Piryatinska et al. \cite{piryatinska2017binary}, 84 Participants (39 Healthy and 45 with Schizophrenia), 16 Channels, 128 Hz \\
		Oh et al. 2019  \cite{Oh2019} & CNN & Schizophrenia & Normal and Schizophrenia  & Private: 28 Participants (14 with Schizophrenia and 14 Normal), 19 Channels, 250 Hz \\
		Phang et al.  2020  \cite{phang_multi-domain_2020}  & CNN & Schizophrenia & Normal and schizophrenia  & Lomonosov Moscow State University Dataset \\				
		\bottomrule

	\end{longtable}
	\clearpage
	\begin{longtable}{p{4cm}p{4cm}p{7cm}p{5cm}}
		\caption{Key Information of Papers about Emotion Recognition} \\
		\label{tab_emotion}\\
		\toprule
		Authors			&	Models	&	Classes	&	Data (Private/Public: No. of Participants, No. of Channels, Sampling Rate)	\\
		\midrule
		Jirayucharoensak et al. 2014  \cite{Jirayucharoensak2014} & SAE & Happy, Pleased, Relaxed, Excited, Neutral, Calm, Distressed, Miserable, and Depressed & DEAP Dataset  \\
		Zheng et al.	2014	\cite{zheng2014eeg}	&	DBN	&	Positive and Negative	&	Private: 6 Participants, 62 Channels, 1000 Hz	\\
		Al-Nafjan et al.  2017  \cite{Al-Nafjan2017}  & DNN   & Excitement, Meditation, Boredom, and Frustration  & DEAP Dataset  \\
		Alhagry et al.  2017  \cite{Alhagry2017}  & LSTM  & High/Low Arousal, High/Low Valence, High/Low Liking & DEAP Dataset  \\
		Li et al. 2017  \cite{Li2017a}  & CNN+LSTM  & High/Low Valence, High/Low Arousal  & DEAP Dataset  \\
		Yin et al.  2017  \cite{Yin2017a} & SAE & High/Low Valence, High/Low Arousal  & DEAP Dataset  \\
		Bozhkov et al.  2017  \cite{Bozhkov2017}  & ESN & Positive and Negative & Private: 26 Participants, 21 Channels, 1000Hz   \\
		Zheng et al.  2017  \cite{zheng_identifying_2017} & ELM & \tabincell{l}{1. High/Low Valence, High/Low Arousal\\
			2. Positive, Neutral, and Negative} & \tabincell{l}{1. DEAP Dataset\\
			2. SEED Dataset}  \\
		Yang et al. 2018  \cite{yang_eeg-based_2018}  & Hierarchical Network  & Positive, Neutral, and Negative & SEED Dataset  \\
		Chen et al. 2018  \cite{Chen2018} & DBN & Happy, Calm, Sad, and Fear  & Private: 10 Participants, 16 Channels, 128Hz  \\
		Hemanth et al.  2018  \cite{Hemanth2018}  & DNN & Happy, Sad, Relaxed, and Angry  & DEAP Dataset  \\
		Choi et al. 2018  \cite{Choi2018} & LSTM  & High/Low Valence, High/Low Arousal  & DEAP Dataset  \\
		Kwon et al. 2018  \cite{Kwon2018} & CNN & High/Low Valence, High/Low Arousal  & DEAP Dataset  \\
		Bagherzadeh et al.  2018  \cite{Bagherzadeh2018}  & SAE & High/Low Valence, High/Low Arousal  & DEAP Dataset  \\
		Chao et al. 2018  \cite{Chao2018} & DBN, RBM  & Pleasant, Unpleasant, Aroused, and Relaxed  & DEAP Dataset  \\
		Li et al. 2018  \cite{Li2018} & CNN & Positive, Neutral, and Negative & SEED Dataset  \\
		Kim et al.  2018  \cite{Kim2018b} & DBN & Relaxed, Fear, Joy and Sad  & Private: 25 Participants, 64 Channels, 1000Hz \\
		Teo et al.  2018  \cite{Teo2018}  & DNN   & \tabincell{l}{1. Like and Dislike\\
			2. Rest and Excited}  & Private: 16 Participants, 9 Channels, - Hz  \\
		Zheng et al.  2019  \cite{Zheng2019}  & RBM, AE & Happy, Sad, Fear, and Neutral   & SEED-IV Dataset \\
		Chao et al. 2019  \cite{Chao2019} & CapsNet & High/Low Arousal, High/Low Valence, High/Low Dominance  & DEAP Dataset  \\
		Chen et al. 2019  \cite{Chen2019} & GRU & High/Low Valence, High/Low Arousal  & DEAP Dataset  \\
		Balan et al.  2019  \cite{Balan2019}  & DNN & No, Low, Medium, and High Fear  & DEAP Dataset  \\
		Zhang et al.  2019  \cite{Zhang2019b} & RNN & Positive, Neutral, and Negtive  & SEED Dataset  \\
		Zeng et al. 2019  \cite{Zeng2019} & CNN & Positive, Neutral, and Negtive  & SEED Dataset  \\
		Gao et al.  2020  \cite{Gao2020}  & CNN & Happy, Sad, and Fear  & Private: 15 Participants, 30 Channels, 1000 Hz  \\
		Serap Aydin 2020  \cite{aydin_deep_2020} & LSTM  & Fear, Anger, Happiness, Sadness, Amusement, Surprise, Excitement, Calmness, Disgust & Private: 23 Participants, 16 Channels, 128 Hz \\
		Cimtay et al. 2020  \cite{cimtay_investigating_2020}  & CNN & \tabincell{l}{1. Positive and Negative\\
			2. Positive, Neutral, and Negative} & \tabincell{l}{1. SEED Dataset\\
			2. DEAP Dataset\\
			3. LUMED Dataset} \\
		Kim et al.  2020  \cite{kim_eeg-based_2020} & CNN+LSTM, LSTM  & \tabincell{l}{1. Low and High\\
			2. Low, Medium, and High} & DEAP Dataset  \\
		Kim et al.  2020  \cite{kim_deep_2018}  & CNN+LSTM  & High/Low Valence, High/Low Arousal  & DEAP Dataset  \\
		Zhu et al.  2020  \cite{zhu_eeg-based_2020} & CNN & Anger, Disgust, Neutral, and Happy  & Private: 40 Participants, 62 Channels, 1000 Hz  \\				
		
		\bottomrule
	\end{longtable}
	\clearpage
	\begin{longtable}{p{2.5cm}p{1.8cm}p{3cm}p{7.5cm}p{6cm}}
		\caption{Key Information of Papers about Operator Functional States} \\
		\label{tab_ofs}\\
		\toprule
		Authors			&	Models	&	Categories	&	Classes	&	Data (Private/Public: No. of Participants, No. of Channels, Sampling
		Rate)	\\
		\midrule
		Chai et al.	2017	\cite{Chai2017}	&	DBN	&	Fatigue	&	Alert and Fatigue 	&	Private: 43 Participants, 32 Channels, 2048 Hz	\\
		Zeng et al.	2018	\cite{Zeng2018}	&	CNN	&	Fatigue	&	Sober and Fatiuge	&	Private: 10 Participants, 16 Channels, 256  Hz	\\
		Yin et al.	2018	\cite{Yin2018}	&	ELM	&	Fatigue	&	Low and High Mental Workload Levels	&	Private: 14 Participants, 11 Channels, 500  Hz	\\
		Ma et al.	2019	\cite{Ma2019}	&	PCANet	&	Fatigue	&	Awake and Fatigue	&	Private: 6 Participants, 32 Channels, 500  Hz	\\
		Gao et al.	2019	\cite{Gao2019}	&	CNN	&	Fatigue	&	Alert and Fatigue	&	Private: 8 Participants, 30 Channels, 1000  Hz	\\
		Jeong et al.	2019	\cite{Jeong2019}	&	CNN+LSTM	&	Mental State and Drowsiness 	&	\tabincell{l}{1. Alert and Drowsy\\
			2. Very Alert, Fairly Alert, neither Alert nor Sleepy, Sleepy but No Effort to Keep Awake, and Very Sleepy}	&	Private: 8 Participants, 30 Channels, 1000 Hz	\\
		Zhang et al.	2017	\cite{Zhang2017}	&	DBN	&	Mental Workload	&	\tabincell{l}{1. Unloaded/Low/Normal/High Level\\
			2. Unloaded/Very/Low/Low/\\Medium/High/Very High/Overloaded Level}	&	Private: 6 Participants, 15 Channels, 500 Hz	\\
		Yin et al.	2017	\cite{Yin2017}	&	SAE	&	Mental Workload	&	Low and High 	&	Private: 7 Participants, 11 Channels, 500  Hz	\\
		Hefron et al.	2018	\cite{Hefron2018}	&	CNN+LSTM	&	Mental Workload	&	Low and High 	&	Private: 8 Participants, 128 Channels, 4096 Hz	\\
		Jiao et al.	2018	\cite{Jiao2018}	&	CNN	&	Mental Workload	&	4 Levels (1, 2, 3, and 4)	&	Private: 13 Participants, 64 Channels, 500  Hz	\\
		Yang et al.	2019	\cite{Yang2019}	&	SAE	&	Mental Workload	&	Low and High 	&	Private: 8 Participants, 11 Channels, 500 Hz	\\
		Tao et al.	2019	\cite{Tao2019}	&	ELM	&	Mental Workload	&	Low and High 	&	Private: 8 Participants, 11 Channels, 500 Hz	\\
		Zhang et al.	2019	\cite{Zhang2019d}	&	CNN+LSTM	&	Mental Workload	&	Low and High 	&	Private: 20 Participants, 16 Channels, 1000 Hz	\\
		Yin et al.	2019	\cite{Yin2019a}	&	DAE	&	Mental Workload	&	Low and High 	&	\tabincell{l}{1. Private: 14 Participants, 11 Channels, 500 Hz\\
			2. DEAP Dataset}	\\
		Zhang et al.	2019	\cite{Zhang2019c}	&	CNN	&	Mental Workload	&	Low, Medium, and High	&	Private: 17 Participants, 16 Channels, 1000 Hz	\\
		Wu et al.	2019	\cite{Wu2019a}	&	CAE	&	Mental Workload and Fatigue	&	Normal, Mild Fatigue, and Excessive Fatigue	&	Private: 40 Participants, 1 Channel, - Hz	\\
		Yin et al.	2017	\cite{Yin2017b}	&	DBN	&	Mental Workload and Fatigue
		&	\tabincell{l}{1. Low, Medium and High Mental Workload \\ 2. Low, Medium and High Fatigue}
		&	Private: 8 Participants, 11 Channels, 500 Hz	\\
		Li et al.	2017	\cite{Li2017}	&	DBN, SAE	&	Mental Workload and Fatigue
		&	Engagement Levels	&	Private: 15 Participants, 32 Channels, 200 Hz	\\
		
		\bottomrule
	\end{longtable}

	\clearpage
	\begin{longtable}{p{3cm}p{3cm}p{3cm}p{5cm}p{9cm}}
		\caption{Key Information of Papers about Sleep Stage Classification } \\
		\label{tab_slepp}\\
		\toprule
		Authors			&	Models	&	Dimension	&	Classes	&	Data (Private/Public: No. of Participants, No. of Channels, Sampling Rate)	\\
		\midrule
		Yildirim et al. 2019  \cite{Yildirim2019} & CNN & EEG, EOG  & W, N1, N2, N3, N4, REM  & \tabincell{l}{1. Sleep-EDF Database\\
			2. Sleep-EDF Database Expanded} \\
		Patanaik et al. 2018  \cite{Patanaik2018} & CNN & EEG, EOG, & W, N1, N2, N3, REM  & Private: Healthy Adolescents and Adults,  Sleep Disorders Patients, Parkinson’s Disease Patients  \\
		Yuan et al. 2019  \cite{Yuan2019} & CNN+GRU & EEG, EOG, EMG   & W, S1, S2, SWS, REM & UCD Database  \\
		Zhang et al.  2019  \cite{Zhang2019g} & CNN+LSTM  & EEG, EOG, EMG   & W, N1, N2, N3, REM  & SHHS  \\
		Chapotot et al. 2010  \cite{Chapotot2010} & MLP & EEG, EOG, EMG   & W, N1, N2, N3, Paradoxical Sleep, and Movement Time & Private: 13 Participants, 4 Channels, 128 Hz  \\
		Malafeev  2018  \cite{Malafeev2018} & LSTM, CNN+LSTM & EEG, EOG, EMG   & W, N1, N2, N3, REM  & \tabincell{l}{Private: 18 Healthy Participants, 12 Channels,  256 Hz\\
			Private: 28 patients with narcolepsy and hypersomnia, 6 Channels,  200 Hz}  \\
		Zhang et al.  2016  \cite{Zhang2016}  & DBN & EEG, EOG, EMG   & W, S1, S2, SWS, REM & UCD Database  \\
		Phan et al. 2019  \cite{Phan2019} & CNN & EEG, EOG, EMG   & W, N1, N2, N3, REM  & \tabincell{l}{1. MASS Database\\
			2. Sleep-EDF Database}  \\
		Chambon et al.  2018  \cite{Chambon2018}  & CNN   & EEG, EOG, EMG       & W, N1, N2, N3, REM  & MASS Database \\
		Jaoude et al. 2020  \cite{jaoude_expert-level_nodate} & CNN+RNN & EEG, EOG, EMG   & W, N1, N2, N3, REM  & \tabincell{l}{1. Private: 6341 Participants, 6 channels, - Hz\\
			2. Private: 93 participants, 6 channels, - Hz\\
			3. From Rosen et al. \cite{rosen_multisite_2012}, 243 patients \\
			4. From Bakker et al. \cite{bakker_gastric_2018},  49 patients} \\
		Biswal et al. 2018  \cite{Biswal2018} & CNN+RNN & EEG, EMG & Sleep Staging, Sleep Apnea, and Limb Movements  & \tabincell{l}{1. SHHS Database\\
			2. From Massachusetts General Hospital Sleep Lab, 10000 Participants, 6 EEG Channels, 200 Hz} \\
		Sors et al. 2018  \cite{Sors2018} & CNN   & Single Channel EEG  & W, N1, N2, N3, REM  & SHHS  \\
		Kulkarni et al. 2019  \cite{Kulkarni2019} & CNN+LSTM  & Single Channel EEG  & Spindles, Non-Spindles in N2 and N3 Stages  & \tabincell{l}{1. MASS Database\\
			2. The DREAMS Sleep Spindles Database\\
			3. From Blank et al. \cite{blank2005overview}, 5 Participants, 2 Channels, 200-512 Hz\\
			4. From Redline et al. \cite{redline2011childhood}, 5 Participants, 2 Channels, 200-512 Hz\\
			5. Private: 18 Epileptic Patients, 1 Channel, 512 Hz} \\
		Tsinalis et al. 2016  \cite{Tsinalis2016} & SAE & Single Channel EEG  & W, N1, N2, N3, REM  & Sleep-EDF Database Expanded \\
		Zhang et al.  2018  \cite{Zhang2018}  & CNN & Single Channel EEG  & W, S1, S2, SWS, REM & \tabincell{l}{1.UCD Database\\
			2.MIT-BIH Polysomnographic Database}  \\
		Mousavi et al.  2019  \cite{Mousavi2019}  & CNN & Single Channel EEG  & W, N1, N2, N3, N4, REM  & Sleep-EDF Database  \\
		Supratak et al. 2017  \cite{Supratak2017} & CNN+LSTM  & Single Channel EEG  & W, N1, N2, N3, REM  & \tabincell{l}{1. MASS Databse\\
			2. Sleep-EDF Database } \\
		Dong et al. 2018  \cite{Dong2018} & LSTM  & Single Channel EEG  & W, N1, N2, N3, REM  & Private:62  Participants, 20 Channels, - Hz \\
		Zhang et al.  2020  \cite{Zhang2020}  & CNN & Single Channel EEG  & W, S1, S2, SWS, REM & \tabincell{l}{1.UCD Database\\
			2. MIT-BIH Polysomnographic Database} \\
		Bresch et al. 2018  \cite{Bresch2018} & CNN+LSTM  & Single Channel EEG  & W, N1, N2, N3, REM  & \tabincell{l}{1. The SIESTA Normative Database\\
			2. Private: 29 Participants, 1 Channel, 1000 Hz}  \\
		AlMeer et al. 2019  \cite{AlMeer2019} & DNN & Single Channel EEG  & W, N1, N2, N3, REM  & Sleep-EDF Database  \\
		Qu et al. 2020  \cite{qu_residual_2020} & CNN & Single channel EEG  & W, N1, N2, N3, REM  & \tabincell{l}{1. MASS Database\\
			2. Sleep-EDF Database}  \\
		Hartmann et al. 2019  \cite{Hartmann2019} & LSTM  & Multiple Channels EEG & Consecutive Activation Phases and Background Phase  & CAP Sleep Database  \\
		Charnbon et al. 2019  \cite{Charnbon2019} & CNN & Multiple Channels EEG & Spindles, K-complexes, and Arousals & \tabincell{l}{1. MASS Database\\
			2. Stanford Sleep Cohort Dataset \cite{andlauer2013nocturnal}, 26 Participants, 1 Channel (C4 or C3), 128 Hz\\
			3. WisConsin Sleep Cohort Dataset \cite{young2008sleep}, 30 Participants, 1 Channel (C4 or C3), 200 Hz\\
			4. MESA}  \\
		Jeon et al. 2019  \cite{Jeon2019} & CNN+LSTM  & Multiple Channels EEG & W, N1, N2 & Private: 218 Pediatric Participants, 32 Channels, 200 Hz  \\
		Chriskos et al. 2020  \cite{Chriskos2020} & CNN & Multiple Channels EEG & N1, N2, N3, REM   & Private: 22 Participants, 19 Channels, - Hz \\				
		\bottomrule
		
	\end{longtable}
	\clearpage
	\begin{longtable}{p{2.5cm}p{1.8cm}p{4cm}p{5cm}p{8cm}}
		\caption{Key Information about Other Applications} \\
		\label{tab_other_applications} \\
		\toprule
		Authors			&	Models	&	Categories	&	Classes	&	Data (Private/Public: No. of Participants, No. of Channels, Sampling
		Rate)	\\
		\midrule
		Kaushik et al.	2019	\cite{Kaushik2019}	&	LSTM	&	Age and Gender Prediction	&	Class from 0 to 5, Varies from Age and Gender	&	From Kaur et al. \cite{kaur2019age}, 60 Participants (35 males and 25 females), 14 Channels, - Hz	\\
		Wulsin et al.	2012	\cite{Wulsin2011}	&	DBN	&	Anomaly Detection	&	5 Classes: Spike and Sharp Wave, GPED  and Triphasic, PLED, Eye Blink, and Background	&	Private: 11 Participants, 17 Channels, 256 Hz	\\
		Anem et al.	2019	\cite{Anem}	&	CNN+LSTM	&	Artifacts Removal	&	-	&	-	\\
		Jacob et al.	2019	\cite{Jacob2019}	&	-	&	Artificial Muscle Intelligence System	&	Grasp, Release, Rollup, Rolldown, and Rollup Release	&	Private: 20 Participants (10 Healthy and 10 Paralyzed), 16 Channels, - Hz	\\
		Huang et al.	2018	\cite{Huang2018}	&	CNN	&	Auditory Salience	&	4923 Classes of Video Classification	&	Private: - Participants, 128 Channels, 2048 Hz	\\
		Yang et al.	2018	\cite{Yang2018}	&	SAE	&	Automatic Ocular Artifacts Removal	&	-	&	BCI Competition IV Dataset 1	\\
		Jiang et al.	2019	\cite{Jiang2019}	&	CNN+LSTM	&	Brain Imaging Classification	&	40 Classes of Images	&	ImageNet-EEG Dataset \cite{spampinato2017deep}	\\
		Baltatzis et al.	2017	\cite{Baltatzis2017}	&	CNN	&	Bullying Incidences Identification	&	Bullying 2D/VR, Non-Bullying 2D/VR	&	Private: 18 Participants, 256 Channels, 250 Hz	\\
		Toraman et al.	2019	\cite{Toraman2019}	&	CNN	&	Cerebral Dominance Detection	&	Left and Right-Hemisphere Dominance	&	Private: 67 Participants (35 Right-Hand Dominat and 32 Left-Hand Dominat), 18 Channels, - Hz	\\
		Doborjeh et al.	2018	\cite{Doborjeh2018}	&	SNN	&	Classification of Familiarity of Marketing Stimuli	&	Familiar and Unfamiliar Brands	&	Private: 20 Participants, 19 Channels, 256 Hz	\\
		Croce et al.	2019	\cite{Croce2019}	&	CNN	&	Classification of Independent Components	&	Brain ICs and Artifact ICs	&	Private: - Participants, 128 Channels, 500 Hz	\\
		Zheng et al.	2020	\cite{Zheng2020}	&	LSTM+CNN, GAN	&	Decoding Human Brain Activity	&	40 Classes of Images	&	From Spampinato et al. \cite{spampinato2017deep}, 6 Participants, 128 Channels, 1000 Hz	\\
		Ming et al.	2019	\cite{Ming2019}	&	SAN	&	EEG Data Analysis	&	\tabincell{l}{1. Different Vigilance Stages\\
			2. P300 and Non-P300}	&	\tabincell{l}{1. Private: - Participants, - Channel, 500 Hz\\
			2. From Wu et al., 18 Participants, 64 Channels, 512 Hz}	\\
		Nagabushanam et al.	2019	\cite{Nagabushanam}	&	LSTM	&	EEG Signal Classification	&	-	&	From Bonn University, - Participants, 20 Channels, - Hz	\\
		Hua et al.	2019	\cite{Hua2019}	&	SAE	&	Functional Brain Network	&	High and Low Proficiency Operators	&	Private: 20 Participants, 8 Channels, 1000 Hz	\\
		Goh et al.	2018	\cite{Goh2018}	&	SSRL	&	Gait Pattern Classification	&	Free Walking, Exoskeleton-Assisted Walking at Zero, Low, and High Assistive Forces	&	Private: 27 Participants, 20 Channels, 1000 Hz	\\
		Fares et al.	2019	\cite{Fares2019}	&	LSTM	&	Image Classification	&	40 Classes of Images	&	From Spampinato et al. \cite{spampinato2017deep},6 Participants, 128 Channels, 1000 Hz	\\
		Akbari et al. 	2019	\cite{Akbari2019}	&	DNN	&	Intelligible Speech Recognition	&	-	&	Private: 5 Participants, - Channel, 3000 Hz 	\\
		Antoniades	2018	\cite{Antoniades2018}	&	CNN	&	Mapping Scalp EEG to iEEG	&	-	&	Private: 18 Participants, 32 Channels (12 FO and 20 Scalp), 200 Hz	\\
		Bird et al.	2019	\cite{Bird2019}	&	MLP, LSTM	&	Optimise the Topology of ANN	&	\tabincell{l}{1. Relaxed, Concentrative, and Neutral\\
			2. Positive, Neutral, and Negative\\
			3. 0 to 9 Imaginary EEG}	&	\tabincell{l}{1. EEG Brainwave Dataset: Mental State\\
			2. EEG Brainwave Dataset: Feeling Emotions\\
			3. MindBigData Dataset}	\\
		Wang et al.	2019	\cite{Wang2019}	&	CNN	&	Person Identification	&	-	&	\tabincell{l}{1. From PhysioNet (missing detail), 109 Participants, 64 Channels, 160 Hz\\
			2. Private: 59 Participants, 46 Channels, 250 Hz}	\\
		Ozdenizci et al.	2019	\cite{Ozdenizci2019}	&	CNN	&	Person Identification	&	-	&	Private: 10 Participants, 16 Channels, 256 Hz	\\
		Singhal et al.	2018	\cite{Singhal2018}	&	DBCS	&	Reconstruction and Analysis of Biomedical Signals	&	-	&	\tabincell{l}{1. From Andrzejak et al. \cite{andrzejak2001indications}, 10 Participants (5h Healthy and 5 Epileptic Patients)\\
			2. BCI Competition II and III}	\\
		Gogna et al.	2017	\cite{Gogna2017}	&	SAE	&	Reconstruction and Analysis of Biomedical Signals	&	-	&	From Andrzejak et al. \cite{andrzejak2001indications}, 10 Participants (5 Healthy and 5 Epileptic Patients)	\\
		Jang et al.	2019	\cite{Jang2019}	&	CNN	&	Seizure Detection of Mice	&	Seizure and Non-Seizure	&	Private: Total 4704h of EEG Recording, 1000 Hz	\\
		Arora et al.	2018	\cite{Arora2018}	&	LSTM	&	Successful Episodic Memory Encoding Prediction	&	Successful and Unsuccessful Recall	&	From UT Southwesetern Medical Center: 30 Participants (15 Dominat and 15 Non-Dominant Hemisphere), 13 and 17 Channel (8-14 Contacts per Electrode), 1000 Hz	\\
		Yu et al.	2020	\cite{Yu2020}	&	CNN	&	Tonic Cold Pain Assessment	&	No Pain, Moderate Pain, and Sever Pain	&	Private: 32 Participants, 32 Channels, 500 Hz	\\
		Ogawa et al.	2018	\cite{Ogawa2018}	&	LSTM	&	Video Classification	&	Liked Video and Not Liked Video	&	Private: 11 Participants, 1 Channel, 1024 Hz	\\
		Said et al.	2018	\cite{BenSaid2018}	&	SAE	&	Vital Signs Compression and Energy Efficient Delivery	&	-	&	DEAP Dataset	\\
		
		\bottomrule
	\end{longtable}
	\clearpage
	
\begin{longtable}[r]{p{4.3cm}p{3.5cm}p{4cm}p{4cm}p{8cm}}	
			\caption{A Summary of Datasets Mentioned in This Survey}\\
			\label{tab_datasets}\\
			\toprule									
			Dataset Name	&	Modality	&	Data Information	&	Category	&	URL		\\
			\midrule
			BCI Challenge & EEG & 26 Participants, 56 Channels, 600 Hz  & P300 and Non-P300 & https://www.kaggle.com/c/inria-bci-challenge    \\
			BCI Competition Data & EEG & Multiple Datasets & Multiple Categories     & http://www.bbci.de/competition    \\
			BNCI Horizon  & EEG & Multiple Datasets & Multiple Categories   & http://bnci-horizon-2020.eu/database/data-sets    \\
			CAP Sleep Database  & EEG, EOG, EMG, ECG  & 16 Participants, 3 EEG Channels & W, S1, S2, S3, S4, and REM  & https://physionet.org/content/capslpdb/1.0.0    \\
			CHB-MIT Scalp EEG Database  & EEG & 22 Participants, 23 Channels, 256 Hz  & Ictal Activity, Siezure Onset, and Ofsset & https://physionet.org/content/chbmit/1.0.0    \\
			CSU BCI Collection  & EEG & Vary with data sets in the database & Normal and Motor Impairments  & https://www.cs.colostate.edu/eeg    \\
			DEAP Dataset  & EEG and Physiological Signals & 32 Participants, 32 Channels, 512 Hz  & Scores For Arousal, Valence, Iiking, Dominance and Familiarity  & http://www.eecs.qmul.ac.uk/mmv/datasets/deap    \\
			EEG Brainwave Dataset: Feeling Emotions & EEG & 2 Participants, 4 Channels, - Hz  & Positive, Neutral,  and Negative  & https://www.kaggle.com/birdy654/eeg-brainwave-dataset-feeling-emotions    \\
			EEG Brainwave Dataset: Mental State & EEG & 4 Participants, 4 Channels, - Hz  & Relaxed, Concentrating, and Neutral & https://www.kaggle.com/birdy654/eeg-brainwave-dataset-mental-state    \\
			EEG Database Data Set/UCI EEG Dataset & EEG & 122 Participants, 64 Channels, 256 Hz & Alcoholic and Control & https://archive.ics.uci.edu/ml/datasets/eeg+database    \\
			EEG Motor Movement/Imagery Dataset
			& EEG & 109 Participants, 64 Channels, 160 Hz & Left/Right Fist or Both Fists/Both Feet & https://physionet.org/content/eegmmidb/1.0.0    \\
			
			EPFL BCI Dataset  & EEG & 9 Participants, 34 Channels, 2047 Hz  & P300 and Non-P300 & https://www.epfl.ch/labs/mmspg/research/page-58317-en-html/bci-2/bci\_datasets/emotion\_dataset/   \\
			Epileptic Seizure Recognition Data Set  & EEG & 500 Participants, - Channels, 173.61 Hz & Healthy With Eyes Open/Closed, Patients during Seizure/Interictal from Hippocampal Location/Interictal from Epileptogenic Zone  & https://archive.ics.uci.edu/ml/datasets/Epileptic+Seizure+Recognition   \\
			MASS Database   & EEG, EOG, EMG, ECG  & 200 Participants,  4–20 EEG Channels, 256Hz & W, N1, N2, N3, and REM  & http://www.ceams-carsm.ca/en/MASS   \\
			MESA  & EEG, EOG  & 6814 Participants, Fz-Cz, Cz-Oz, C4, 256Hz  & Arousal Level & https://www.sleepdata.org/datasets/mesa   \\
			MindBigData & EEG & Vary with data sets in the dataset  & Brain Reaction from Seeing A Digit (0 to 9) & http://www.mindbigdata.com/opendb   \\
			MIT-BIH Polysomnographic Database & EEG, ECG, EOG, EMG, Respiration Signals, and Physiological Signals  & 60 subjects, 7 PSG Channels, 250 Hz & W, N1, N2, N3, N4, and REM  & https://www.physionet.org/content/slpdb/1.0.0   \\
			SEED Dataset  & EEG and  Eye Movement & 15 Participants, 62 Channels, 1000Hz  & \tabincell{l}{Positive/Neutral/Negative \\and Happy/Sad/Neutral/Fear}  &  https://bcmi.sjtu.edu.cn/home/seed/   \\
			SHHS  & EEG, EOG, EMG & 6,441 Participants, C4-A1 and C3-A2, 125 Hz & W, N1, N2, N3, N4, and REM  & https://sleepdata.org/datasets/shhs/    \\
			Sleep-EDF Database Expanded & EEG, EOG, EMG & 61 Participants, Fpz-Cz and Pz-Oz, 100 Hz & W, S1, S2, S3, S4, and REM  & https://physionet.org/content/sleep-edfx/1.0.0    \\
			Sleep-EDF Database
			
			& EEG, EOG, EMG & 20 Participants, Fpz-Cz and Pz-Oz, 100 Hz & W, N1, N2, N3, and REM  & https://physionet.org/content/sleep-edf/1.0.0   \\
			The Bern-Barcelona EEG Database & EEG & 5 Participants, 7500 Pairs of Signals,  512 or 1024 Hz  & Focal and Non-Focal & https://www.upf.edu/web/mdm-dtic/datasets   \\
			
			The SIESTA Normative Database (cross-institute) & EEG, EOG, EMG, ECG  & 292 Participants, 6 EEG Channels,  Variable (minimum 100Hz) & W, N1, N2, N3, and REM  & http://ofai.at/siesta/database.html   \\
			
			UCD Database  & EEG and Physiological Signals & 25 Participants, C3–A2 and C4–A1, 128Hz & W, S1, S2, Sws, and REM & https://physionet.org/content/ucddb/1.0.0   \\
			LUMED Dataset & EEG and Physiological Signals & 11 Participants, 8 Channels, 500 Hz & Negative and Positive Valence & https://www.dropbox.com/s/xlh2orv6mgweehq/LUMED\_EEG.zip?dl=0    \\
			
			\bottomrule

		\end{longtable}
	\end{spacing}

\end{landscape}
\begin{table*} 
	\caption{A Brief Summary of Sleep Stages}
	\label{tab_sleep}
	\centering
	\begin{tabular}{lll}  
		\toprule   	
		Sleep Stages	&	Main Features of EEG in Each Stage	&	Brief Description	\\
		\midrule
		Wake	&	Alpha Waves	&	Before Sleep	\\
		Stage N1 NREM	&	Low-Voltage Theta Waves	&	Blood Pressure Falls	\\
		Stage N2 NREM	&	Theta Waves with K Complexes and Sleep Spindles	&	Cardiac Activity Decrease	\\
		Stage N3 NREM	&	High-Amplitude Delta Waves	&	High Threshold for Arousal	\\
		Stage REM Sleep	&	Low-Amplitude Theta Waves	&	Blood Pressure and Pulse
		Rate Increase	\\
		\bottomrule  
	\end{tabular}
\end{table*}

\end{document}